\documentclass[onecolumn,showpacs,preprintnumbers,nofootinbib,amsmath,amssymb]{revtex4}

\usepackage{amsfonts} 
\usepackage{amssymb}  
\usepackage{graphicx} 
\usepackage{amsmath}  
\usepackage{amsmath}  
\usepackage[latin1]{inputenc}
\usepackage{color}

\begin{document}
\title{(Self-)Magnetized Bose-Einstein Condensate stars}
\author{G. Quintero Angulo\footnote{gquintero@fisica.uh.cu}}
\affiliation{Facultad de F{\'i}sica, Universidad de la Habana,\\ San L{\'a}zaro y L, Vedado, La Habana 10400, Cuba}
\author{A. P\'erez Mart\'{\i}nez\footnote{aurora@icimaf.cu}}
\affiliation{Instituto de Cibern\'{e}tica, Matem\'{a}tica y F\'{\i}sica (ICIMAF), \\
 Calle E esq a 15 Vedado 10400 La Habana Cuba\\}
\author{H. P\'erez Rojas\footnote{hugo@icimaf.cu}}
\affiliation {Instituto de Cibern\'{e}tica, Matem\'{a}tica y F\'{\i}sica (ICIMAF), \\
 Calle E esq a 15 Vedado 10400 La Habana Cuba\\}
 \author{D. Manreza Paret\footnote{dmanreza@fisica.uh.cu}}
\affiliation {Facultad de F\'{\i}sica, Universidad de La Habana, \\ San L{\'a}zaro y L, Vedado, La Habana 10400, Cuba.\\
Instituto de Ciencias Nucleares, Universidad Nacional Aut\'{o}noma de M\'{e}xico, Apartado Postal 70-543, CdMx 04510, M\'{e}xico\\}
\thanks{}%
\date{\today}%
\begin{abstract}

We study magnetic field effects on the Equations of State (EoS) and the structure of Bose-Einstein Condensate (BEC) stars, i.e. a compact object composed by a gas of interacting spin one bosons formed up by the pairing of two neutrons. To include the magnetic field in the thermodynamic description, we suppose that particle-magnetic field and particle-particle interactions are independent.
We consider two configurations for the magnetic field: one where it  constant and externally fixed, and another where it is produced by the bosons by self-magnetization. Since the magnetic field produces the splitting of pressures in the directions along and perpendicular to the magnetic axis, stable configurations of self-magnetized and magnetized BEC stars are studied using the recently found $\gamma$-structure equations that describe axially symmetric objects. The magnetized BEC stars are, in general spheroidal, less massive and smaller than the non-magnetic ones, being these effects more relevant at low densities. For the self-magnetized BEC stars their inner profiles of magnetic field can be computed as a function of the equatorial radii. The values obtained for the core and surface magnetic fields are in agreement with those typical of compact objects.

\end{abstract}


\maketitle

\section{Introduction}	

Cold fermion gases describe with relative success compact objects like white dwarfs (WD) and neutron stars (NS) \cite{Shapiro,camenzind2007compact}. The stable configurations -masses and radii- of these compact objects are obtained by the combination of the Equations of State\footnote{The pressure and the energy density as a parametric function of the particle density/chemical potential.} (EoS) of the matter that constitutes them, and the so called structure equations, which guarantees hydrodynamic equilibrium in the context of General Relativity \cite{Shapiro}. However, the problem of an accurate description of compact objects is far from being exhausted, specially in what concerns neutron stars, mainly due to the uncertainties about their composition and the behavior of matter at supra-nuclear densities \cite{camenzind2007compact}.

Many kinds of exotic matter and phases have been conjectured to exist in NS interior \cite{camenzind2007compact}, being one of the most interesting the possibility of nucleon pairing \cite{Sauls:1989}. Although the idea of neutron stars having a superfluid crust and core is more than thirty years old, until now the strongest observational argument to support it comes form the adjustment of the cooling data of the NS of Cassiopeia A \cite{Page:2011yz,PageSC,Shternin:2010qi}. Up to this time, it is believed that neutrons in the crust paired in a spin antiparallel state, while in the core they paired with parallel spins giving birth to magnetic Cooper pairs. Protons in the core pair in a spin antiparallel state. The critical temperatures for each of these process to occur are in a range of $T_c \simeq 10^8-10^{10}$~K, being this order of temperatures reached in neutron stars a few hundred years after they birth \cite{PageSC,Shternin:2010qi}. Since the transition to the superfluid phase is of second order, unless the temperature of the star is $T \ll T_c$ in which case the nucleon gases are almost fully paired, the pairing and thermal unpairing processes are balanced, and a mixture of Cooper pairs and free fermions will coexist in the interior of the NS crust and core \cite{PageSC}.

After the enormous advances made from the $2000$'s in the cooling of atomic systems, it has been possible to prove experimentally that superfluidity and Bose-Einstein condensation (BEC) are not less that extreme states of the same phenomenon of fermion pairing \cite{Randeria,Leggett,Parish}. In the superfluid -BCS- limit, fermions form weakly bounded Cooper pairs whose size is of the order of the interparticle distance. On the contrary, in the BEC limit, fermions are tightly bounded forming dimers molecules that behave as bosons. The critical temperature for pairing $T_p$, that coincides with the critical temperature for superfluidity in the BCS limit, has been estimated to be highest far from the condensation temperature ($T_c$) in the BEC limit. This means that the fermion gas might be bosonized although not condensed. In regard of the experimental achievement of the BCS-BEC crossover, the possibility that at some stage of its evolution neutron star cores are fully or mainly composed by a gas of bosons -paired fermions- can not be discarded a priori.

Theoretical modeling of stars fully composed by bosons goes back to the decade of the sixties of the last century \cite{Ruffini,Takasugi,Gleiser}. A boson gas at low temperatures undergoes a phase transition to the Bose-Einstein condensate \cite{Landau}. Therefore, in the most idealized case of a star formed by non interacting bosons at zero temperature, the gas does not exert pressure and the usual scheme of EoS plus structure equations cannot be used to find the mass and radius of these stars. Albeit, it is still possible to obtain them in the frame of self-consistent field methods \cite{Ruffini}. For non interacting boson stars composed by particles with masses in the order of nucleon mass, the star masses are about $10^{-20}M_{\odot}$ \cite{Takasugi}, a discouraging result when your aim is to describe a compact object. Yet, more massive boson stars might be obtained by adding repulsive self-interacting terms to the models \cite{Grandclement}.

Notwithstanding, boson stars have witnessed a revival due to the experimental findings previously mentioned: the BEC for non ideal gases of composite bosons \cite{Anderson:1995gf}, the observation in lab of the BEC-BCS crossover and the success in the adjustment of the cooling data of the neutron star of Cassiopeia A. Within these new models, known as Bose Einstein condensate (BEC) stars, interaction is key to counterbalance gravity allowing to formulate a well defined EoS for the boson gas that combined with structure equations yield stable configurations of masses and radii \cite{Chavanis,Latifah}. In \cite{Chavanis} polytropic EoS are obtained for no relativistic (Gross-Pitaviskii with short range Van der Waals type repulsion) and relativistic (Klein-Gordon with $\lambda \phi^4$ interaction) stars of scalar boson gases at zero temperature. In \cite{Latifah}, to obtain the EoS, a Van der Waals repulsive interaction is added to the free particle non relativistic Hamiltonian and temperature is included. In both cases, the BEC stars are taken as alternative description of neutron stars interiors in which boson are formed by the pairing of two neutrons. The parameters governing the size and weight of the resulting stars are the boson mass and the strength of the interaction. By an appropriate selection of this values, stellar masses around two solar masses can be reached \cite{Chavanis}.

However, none of these works take into account the huge magnetic fields present in most neutron stars, whose observed surface values goes from $10^{12}-10^{13}$~G for pulsars to $10^{15}$~G for magnetars \cite{Malheiro:2013loa}, while in their interiors might be up to $10^{18}$~G \cite{duncan,1991ApJ...383..745L,Lattimerprognosis}. This is of particular relevance since, as was mention before, in the core of neutron stars, neutrons pair in a spin parallel state giving rise to a neutral magnetic -vector- boson.

The thermodynamic properties of a gas of neutral vector bosons interacting with a magnetic field have been studied in \cite{Yamada,Lismary} for non relativistic and in \cite{Quintero2017IJMP,Quintero2017PRC,Quintero2017AN} for relativistic bosons. In general terms, for a magnetized neutral vector boson gas, Bose-Einstein condensation can be reached not only by decreasing the temperature or increasing the particle density, but also by augmenting the magnetic field. The magnetic field presence also breaks the the $SO3$ rotational symmetry provoking the separation of the pressures in two components, one parallel and the other perpendicular to the magnetic axis giving rise to anisotropic EoS, just as occurs for magnetized  fermion gases \cite{Aurora2003EPJC,PhysRevC.77.015807,Ferrer1,Ferrer2010wz}. Depending on the temperature, the particle density and the magnetic field, the lower pressure might be negative pointing out that the system is susceptible to suffer a quantum magnetic collapse \cite{Chaichian1999gd}. Another relevant feature of the magnetized boson gas is that for low enough temperatures the gas spontaneously magnetize. This phenomenon has been called Bose-Einstein ferromagnetism and might be directly linked with magnetic field production inside the star because for astrophysical energy densities the values of the self-generated magnetic field are in the order of those expected in NS \cite{Yamada,Elizabeth,Quintero2017PRC,Lismary}.

All those non trivial magnetic field effects in the boson gas EoS influences the mass, size and shape of the magnetized BEC stars. In this spirit, the goal of this work is to study the magnetic field influence in the structure of BEC stars. To do so, we follow \cite{Latifah} in the sense that we obtain the Equation of State of the system starting from a Hamiltonian that is the sum of two terms, one corresponding to the free particles interacting with the magnetic field  and another describing a Van der Waals repulsive type interaction between particles. The neutral vector bosons that form the system are supposed to be two paired neutrons and to have a mass $m = 2 m_n$ ($m_n$ is the neutron mass), and a magnetic moment $\kappa = 2 \mu_n$ ($\mu_n$ is the neutron magnetic moment).

Since the magnetized compact object EoS are, at least in principle, anisotropic, we are forced to go beyond spherical symmetry to obtain static stable configurations of magnetized BEC stars. Therefore, instead of solving the standard TOV equations used for spherical compact objects, we use the axially symmetric structure equations presented in \cite{Samantha}, that allows to compute mass and radii for axially symmetric objects provided they are spheroidal. Non isotropic structure equations give the chance of quantify the deformation caused in the star by the presence of the magnetic field.

In our study, we consider two magnetic field configurations: one where it is constant along the star and another where it is produced by the bosons through the self-magnetization of the gas. The constant magnetic field approximation is a reasonable assumption to consider magnetic field influence and have been used by us in previous works \cite{Paret2014,Paret2015}. On the other hand, taken the magnetic field as produced by the bosons allows to compute its profiles inside the star, and therefore to evaluate whether spin-one boson self-magnetization is a valid mechanism for astrophysical magnetic field production.

The paper is organized as follows. In Section II we present the EoS for (self-)magnetized BEC stars; relativistic and non relativistic bosons are considered. Section III is devoted to the numerical solutions of the $\gamma$ structure equations for constant and self-generated magnetic field; masses and radii, deformation and magnetic field profiles are discussed. Concluding remarks are given in Section IV.

\section{Equations of State of a magnetized BEC star}

The Hamiltonian of an interacting boson gas can be written as a sum of an ideal gas Hamiltonian $\hat{H}_B$, where $B$ denotes the bosons interacting with the magnetic field (through the magnetic moment)  plus the particle-particle interaction Hamiltonian $\hat{H}_{int}$

\begin{equation}\label{hamiltonian}
\hat{H} = \hat{H}_{B} + \hat{H}_{int}.
\end{equation}

In a first approximation, boson-boson interaction might be taken as a two-body contact interaction $u_0 \delta(r-\bar{r})$, where $r$ and $\bar{r}$ are the positions of the interacting particles \cite{Latifah}. The parameter $u_0 = 4 \pi a/m$ indicates the strength of the interaction; $m$ is the boson mass and $a=1$~fm the scattering length \cite{Chavanis,Latifah}. Under this supposition, $\hat{H}_{int}$ can be  written as

\begin{equation}\label{hamiltonian_int}
\hat{H}_{int} = \frac{1}{2} u_0 \sum_{k,k^{\prime}} \hat{n}_k \hat{n}_k^{\prime},
\end{equation}

\noindent being $\hat{n}_k$ the occupation number operator in state $k$. Since the number of particles in astrophysical scenarios is very large while the temperatures are relatively low, it is to expect quantum fluctuations to be not very significant, and the interaction Hamiltonian can be approximated by its expectation value

\begin{equation}\label{hamiltonian_int1}
\hat{H}_{int} \approx \langle \hat{H}_{int} \rangle = \frac{1}{2} u_0 N^2,
\end{equation}

\noindent where $N = \sum_{k} \langle \hat{n}_k \rangle$ is the mean particle density.

Although this type of particle interaction is very simple, it allows the authors of \cite{Latifah} to obtain thermodynamically consistent EoS for a gas of interacting bosons at finite temperature whose corresponding mass-radius curves has maximum values and shapes that are consistent with other BEC star models \cite{Chavanis}. For that reason, we use it when including the magnetic field effects, but having in mind that it is susceptible of theoretical and numerical improvements, but always very useful as starting and comparison point, as well as for an initial understanding of what the magnetic field might cause.

Using Eqs.~(\ref{hamiltonian})-(\ref{hamiltonian_int1}), one can write the system partition function as

\begin{equation}
\Xi = e^{-\beta V \frac{1}{2} u_0 N^2} \Xi_{B},
\end{equation}

\noindent with $\beta$ the inverse of the absolute temperature $T$ and $\Xi_{B}$ the partition function of the non interacting boson system in presence of the magnetic field (which is computed starting from $H_{B}$). The thermodynamical potential -per unit volume- then reads

\begin{equation}\label{termo}
\Omega(N,B,T) =-\frac{1}{\beta V} \ln \Xi = \frac{1}{2} u_0 N^2 + \Omega_B(N,B,T),
\end{equation}

\noindent $\Omega_{B}$ is the thermodynamic potential of the gas of ideal bosons in the presence of the magnetic field.

As can be seeing from Eq.~(\ref{termo}), the boson-boson interaction term $1/2 u_0 N^2$ enters in the thermodynamical potential, and consequently in the EoS, independently of the non interacting particles/ideal gas part $\Omega_{B}$. In consequence, magnetic field effects will enter in the EoS only through the non interacting gas term $\Omega_{B} = \Omega_{B}(T,N,B)$, with $B$ the magnetic field intensity. Once we find $\Omega_{B}$, the EoS of the magnetized boson system are calculated as \cite{Latifah,Ferrer2010wz}

\begin{subequations}\label{EoS}
\begin{align}
P_{\parallel}&= -\Omega + N \left(\frac{\partial \Omega}{\partial N}\right)_{\mu,T,B} = \frac{1}{2}u_0 N^2 -\Omega_{B}, \label{ppar} \\
	\nonumber\\
	P_{\perp}& = -\Omega + N \left(\frac{\partial \Omega}{\partial N}\right)_{\mu,T,B} + B \left(\frac{\partial \Omega}{\partial B}\right)_{\mu,T}  = \frac{1}{2}u_0 N^2 -\Omega_{B} - B \mathcal M, \label{pper} \\
	\nonumber\\
	E &= \Omega + \mu N - T \left(\frac{\partial \Omega}{\partial T}\right)_{\mu,B} = \frac{1}{2}u_0 N^2 + \Omega_{B} + \mu N - T \left(\frac{\partial \Omega_{B}}{\partial T}\right)_{\mu,B}, \label{energy}\\
	\nonumber\\
	N &= - \left(\frac{\partial \Omega}{\partial \mu}\right)_{T,B} = - \left(\frac{\partial \Omega_{B}}{\partial \mu}\right)_{T,B}, \quad	\mathcal M = - \left(\frac{\partial \Omega_{B}}{\partial B}\right)_{\mu,T},
	\end{align}
\end{subequations}

\noindent $P_{\parallel}$ and $P_{\perp}$ are the pressures parallel and perpendicular to the magnetic field direction, $E$ is the internal energy density, $\mu$ the chemical potential of the bosons and $\mathcal M$ the magnetization .

In Eqs.~(\ref{EoS}), both pressures, $P_{\parallel}$ and $P_{\perp}$  have the contribution of interaction \cite{Latifah}, as well as the pressures produced by the bosons in presence of the magnetic field \cite{Ferrer2010wz}. Besides, the perpendicular pressure  has the anisotropic term $-\mathcal MB$ term induced by the magnetic field. The internal energy $E$ of the gas  also contains the energy density of the interaction.

To compute $\Omega_{B}$ we start from

\begin{equation}\label{termodef}
	\Omega_{B}=\sum_{s=-1,0,1} \frac{1}{\beta}\left(\int\frac{d^3p}{(2\pi)^2} \ln \left( (1-e^{-(\varepsilon(p, B,s)-\mu)\beta})(1-e^{-(\varepsilon(p, B,s)+ \mu)\beta})\right)  + \varepsilon(p,B,s) \right ),
\end{equation}

\noindent provided we know the particle spectrum $\varepsilon(p,B,s)$, where $p$ is the total momentum and $s=-1,0,1$ the spin eigenvalues. The thermodynamical potential Eq.~(\ref{termodef}) has two terms. The first one comes from the integral of the logarithm, depends on $B$, $N$ and $T$ and it is known as the statistical term $\Omega^{st}_{B}$; the second one comes from the integral of the particle spectrum and since it is only $B$ dependent it is known as the vacuum term $\Omega^{vac}_{B}$.

$\Omega_{B}$ is calculated in two cases: the relativistic and the non relativistic limit. At first sight, the relativistic case is  most appropriated due to the highest densities and energies that are expected to exist in neutrons stars, and if we decided to consider the  non-relativistic case was because it is the natural extension of the result for a non magnetic BEC star presented in \cite{Latifah}. Nevertheless, in the rest of this section we will see, that for the values of magnetic field supposed in the interior of neutron stars, the differences between the relativistic and the non relativistic limits are negligible.

The spectrum for a relativistic neutral vector boson gas that interacts with a constant magnetic field directed in the $z$ direction $\textbf{B} = (0,0,B)$ and its non relativistic limit are respectively \cite{Quintero2017IJMP,Yamada}

\begin{subequations}\label{spectrum}
\begin{align}
\varepsilon^{r}(p_z,p_{\perp},B)&= \sqrt{m^2+p_z^2+p_{\perp}^2-2\kappa s B\sqrt{p_{\perp}^2+m^2}},\\
\varepsilon^{nr}(p,B)&=m +p^2/2 m - \kappa s B,
\end{align}
\end{subequations}

\noindent where $p_z$ is the momentum component along the magnetic field, $p_{\perp}$ is the momentum component perpendicular to it, and the supra-index $r$ and $nr$ stand for the relativistic and the non relativistic cases respectively. Computing the integrals and sum of Eq.~(\ref{termodef}) with the use of Eqs.(\ref{spectrum}), the following expressions for the relativistic \cite{Quintero2017PRC} and the non relativistic \cite{Lismary} thermodynamical potential are obtained

\begin{subequations}\label{omegas}
\begin{align}
\Omega_{B}^r&=-\frac{(\varepsilon)^{3/2}}{2^{1/2} \pi^{5/2} \beta^{5/2} (2-b)} Li_{5/2}(e^{\beta \mu^{\prime}})\nonumber\\
+&\frac{m^4}{288 \pi} \left \{ b^2(66-5 b^2)-3(6-2b-b^2) (1-b)^2 \log(1-b)-3(6+2b-b^2)(1+b)^2 \log(1+b) \right \},\label{Omegar}\\
\Omega^{nr}_{B}& = -\frac{m^{3/2}}{(2 \pi)^{3/2} \beta^{5/2}} \{ Li_{5/2}(z_{-})+  Li_{5/2}(z) + Li_{5/2}(z_{+})\},\label{Omeganr}
\end{align}
\end{subequations}

\noindent $ Li_{n}(x)$ is the polylogarithmic function of order $n$.

In Eq.~(\ref{Omegar}), \mbox{$\varepsilon=\sqrt{m^2 - 2 m \kappa B} = m \sqrt{1-b}$} is the ground state energy of the relativistic gas,  $\mu^{\prime} = \mu - \varepsilon$ and $b=B/B_c$ where $B_c = m / 2 \kappa \simeq 10^{20}$~G is the value of the magnetic field for which $\varepsilon=0$, i.e at which the magnetic energy becomes comparable with the rest energy of the particles. In Eq.~(\ref{Omeganr}),  $z=e^{\beta \mu}$ is the fugacity and $z_{\pm} =e^{\beta (\mu \pm \kappa B)}$. The terms that contain the polylogarithms in Eqs.~(\ref{omegas}) are the statistical ones, while the one that only depends on $b$ in Eq.~(\ref{Omegar}) corresponds to the vacuum. Vacuum terms are absent in Eq.~(\ref{Omeganr}) because the result of computing  $\int dp_z\varepsilon(p,B)$ with the non relativistic spectrum is zero.

Although the thermodynamical potentials Eqs.~(\ref{omegas}) allow us to work at finite temperature, we focus our paper in the zero temperature case in order to study exclusively the effects of the magnetic field, while magnetized BEC stars at finite temperature will be tackled in a future work. At $T=0$, the statistical terms are zero for both the relativistic and the non relativistic particles. Therefore $\Omega^{nr} =0$ and $\Omega^{r} =\Omega_{vac}$, where we are calling $\Omega_{vac}$ the second term of Eq.~(\ref{Omegar})

\begin{equation}
\Omega_{vac}=\frac{m^4}{288 \pi} \left \{ b^2(66-5 b^2)-3(6-2b-b^2) (1-b)^2 \log(1-b)-3(6+2b-b^2)(1+b)^2 \log(1+b) \right \}.
\end{equation}

\noindent Combining these limits with Eqs.~(\ref{EoS}), we finally obtain the EoS for the magnetized gas of interacting bosons in the relativistic case

\begin{subequations}\label{EoSRtotal}
	\begin{align}
	P_{\parallel}^r = \frac{1}{2}u_0 N^2 -\Omega_{vac}, \quad\quad &P_{\perp}^r = \frac{1}{2}u_0 N^2-\Omega_{vac}-B {\mathcal M}^r, \quad \quad {\mathcal M}^r = \frac{\kappa}{\sqrt{1-b}} N, \label{EoSRtotal1}\\
	E^r&=\frac{1}{2}u_0 N^2 +m \sqrt{1-b} N + \Omega_{vac},  \label{EoSRtotal3}
	\end{align}
\end{subequations}

\noindent and the non relativistic case

\begin{subequations}\label{EoSNRtotal}
\begin{align}
P_{\parallel}^{nr}= \frac{1}{2}u_0 N^2, \quad\quad &P_{\perp}^{nr}=\frac{1}{2}u_0 N^2-B{\mathcal M}^{nr}, \quad \quad {\mathcal M}^{nr} = \kappa N,\\
E^{nr}& =  \frac{1}{2} u_0 N^2 + (m-\kappa B) N.  \label{EoSNRtotal3}
\end{align}
\end{subequations}

\noindent Note that Eqs.~(\ref{EoSRtotal})-(\ref{EoSNRtotal}) are parametrized with respect to the particle density $N$ and the magnetic field intensity $B$.

Modifications due to the presence of the magnetic field  (with respect to the $B=0$ case) enters in Eqs.~(\ref{EoSRtotal}) in three ways. Firstly, by the magnetic pressure $-B {\mathcal M}^r$  which is always positive and diminishes the perpendicular pressure with respect to the parallel one. Secondly, through the vacuum pressure $-\Omega_{vac}$ which is added to the pressures and subtracted to the energy, and finally by the term $m \sqrt{1-b} N$ that appears instead of $m$ and express the change on the ground state produced by the magnetic field. On the other hand, the non relativistic EoS Eqs.~(\ref{EoSNRtotal}) are modified also through the subtractive term $-B {\mathcal M}^{nr}$ in the perpendicular pressure, and through the decreasing of the energy density now by the substitution of the bosons ground state energy $m$ by $(m-\kappa B)$. To complete the EoS of the magnetized compact object the energy density and pressures of the classical electromagnetic field have to be taken into account. To do so, the so called  Maxwell contribution, $B^2/8 \pi$ has to be added to $E$ and $P_{\perp}$ and subtracted from $P_{\parallel}$ in Eqs.~(\ref{EoSRtotal}) and (\ref{EoSNRtotal}). The actual relevance of each of the modifications the magnetic field introduces in the EoS for the values of magnetic field and particle density here considered are not obvious and will be analyzed below for some characteristic values of $N$ and $B$, with and without the Maxwell contribution.

\subsection{EoS: Numerical results and discussions}
\label{sec4}

We devoted this section to study numerically Eqs.~(\ref{EoSRtotal}) and (\ref{EoSNRtotal}), and in particular the anisotropic pressures behavior with the boson mass density and two different configurations of the magnetic field. In the first one, the magnetic field is constant and externally fixed. In the second, it is produced by the bosons through self-magnetization. With this we aim to give a more realistic description of the star inner magnetic field by proposing a mechanism to produce it that is directly related to the matter that composes the compact object.

\subsubsection{EoS with constant magnetic field}
\label{sec4A}

\begin{figure}[h]
	\centering
	\includegraphics[width=0.42\linewidth]{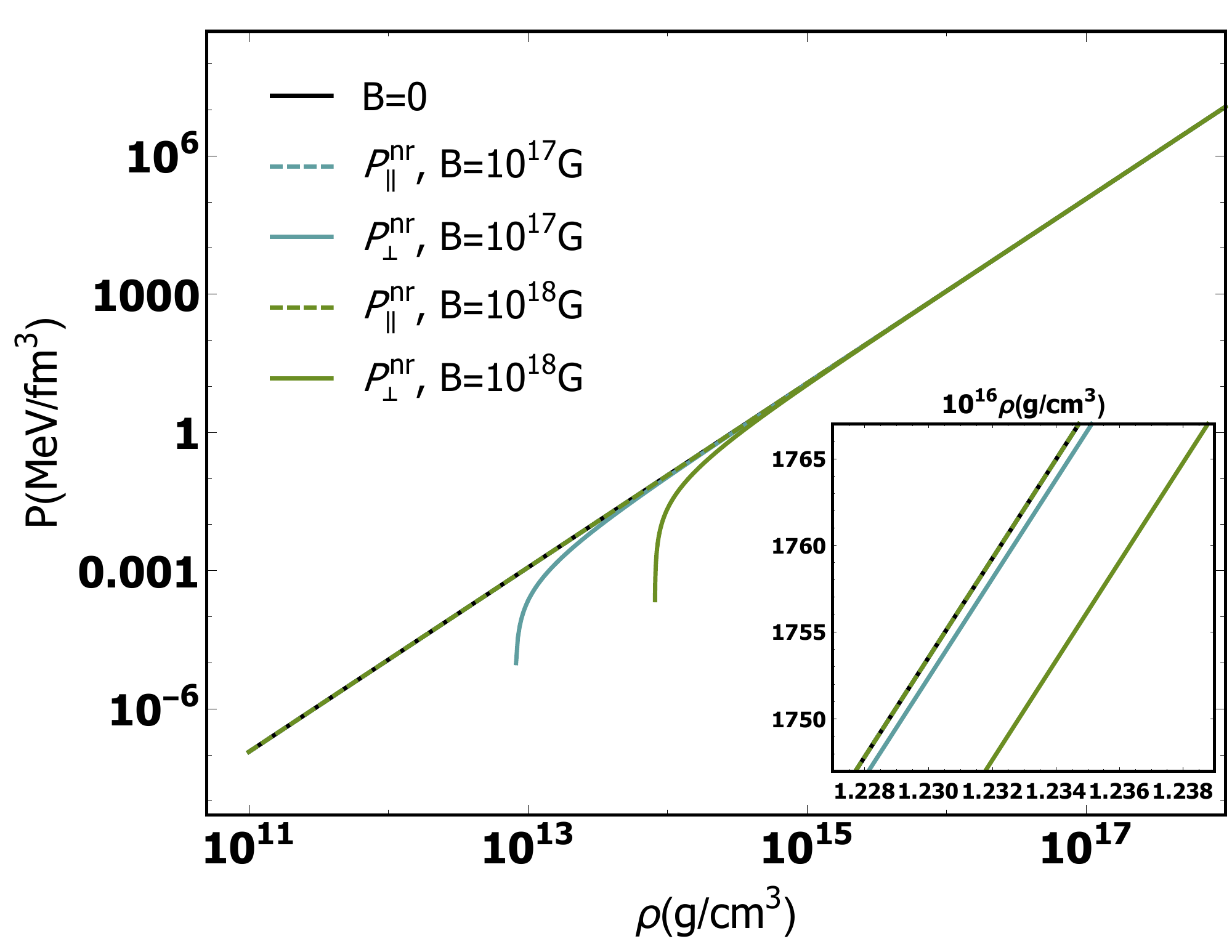}
	\includegraphics[width=0.42\linewidth]{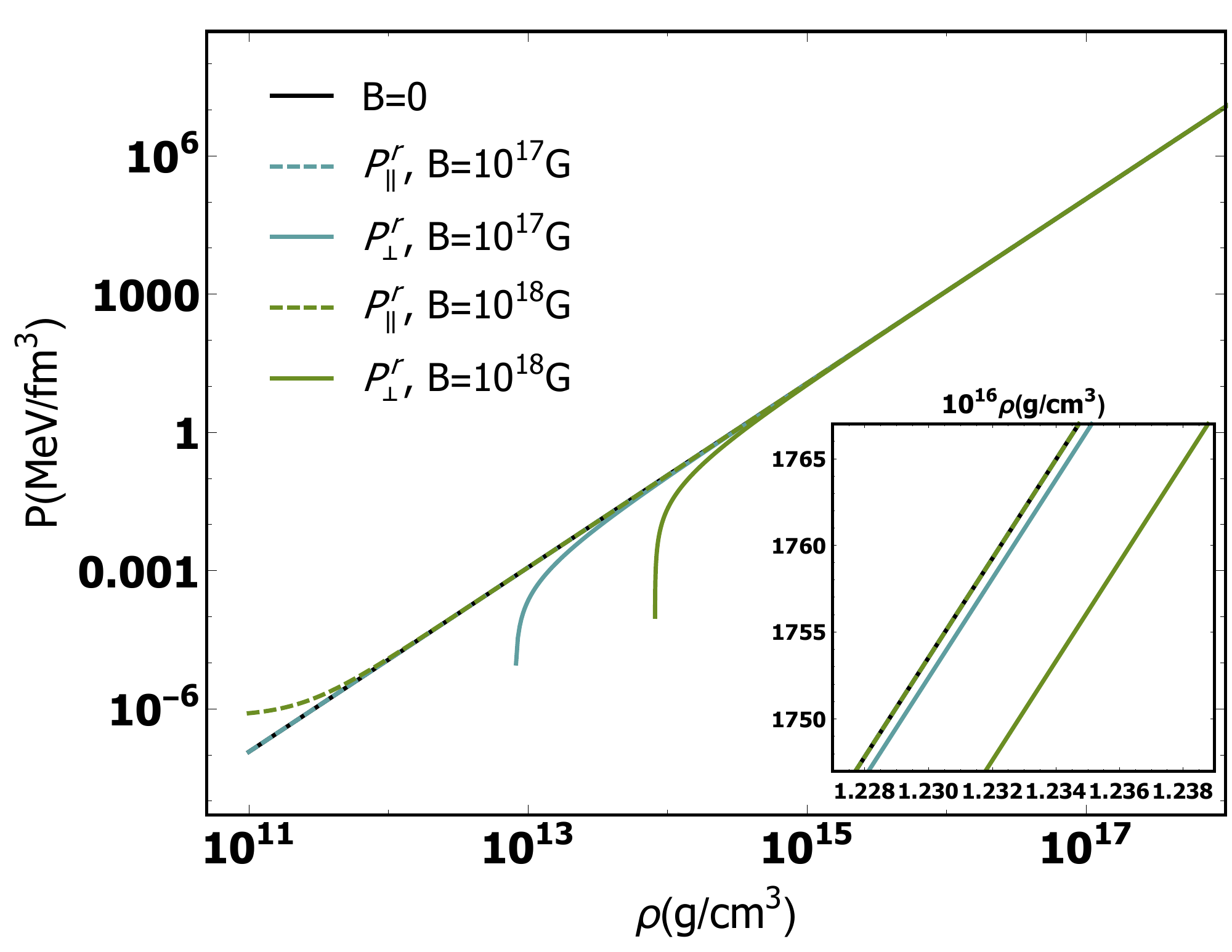}\\
	\includegraphics[width=0.42\linewidth]{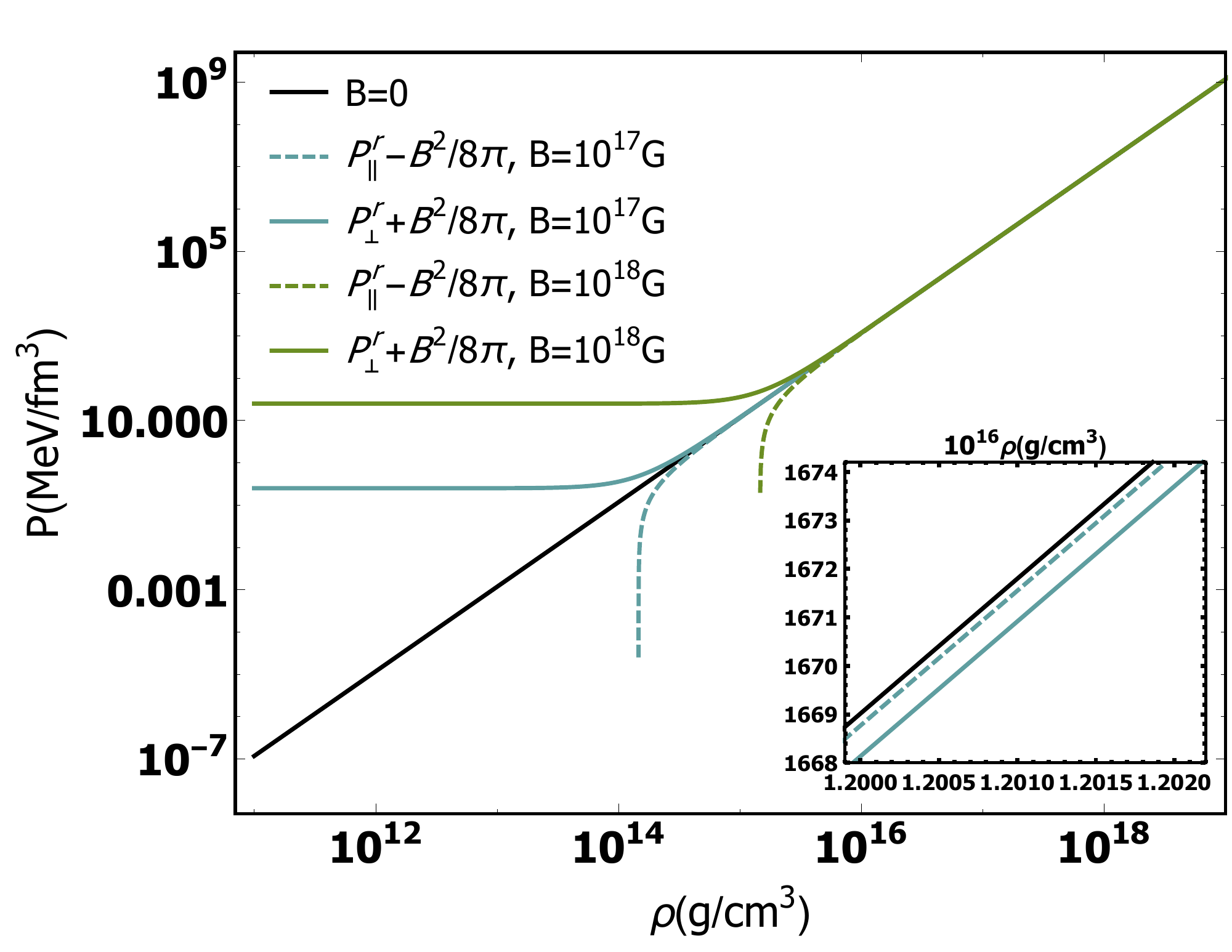}
	\includegraphics[width=0.42\linewidth]{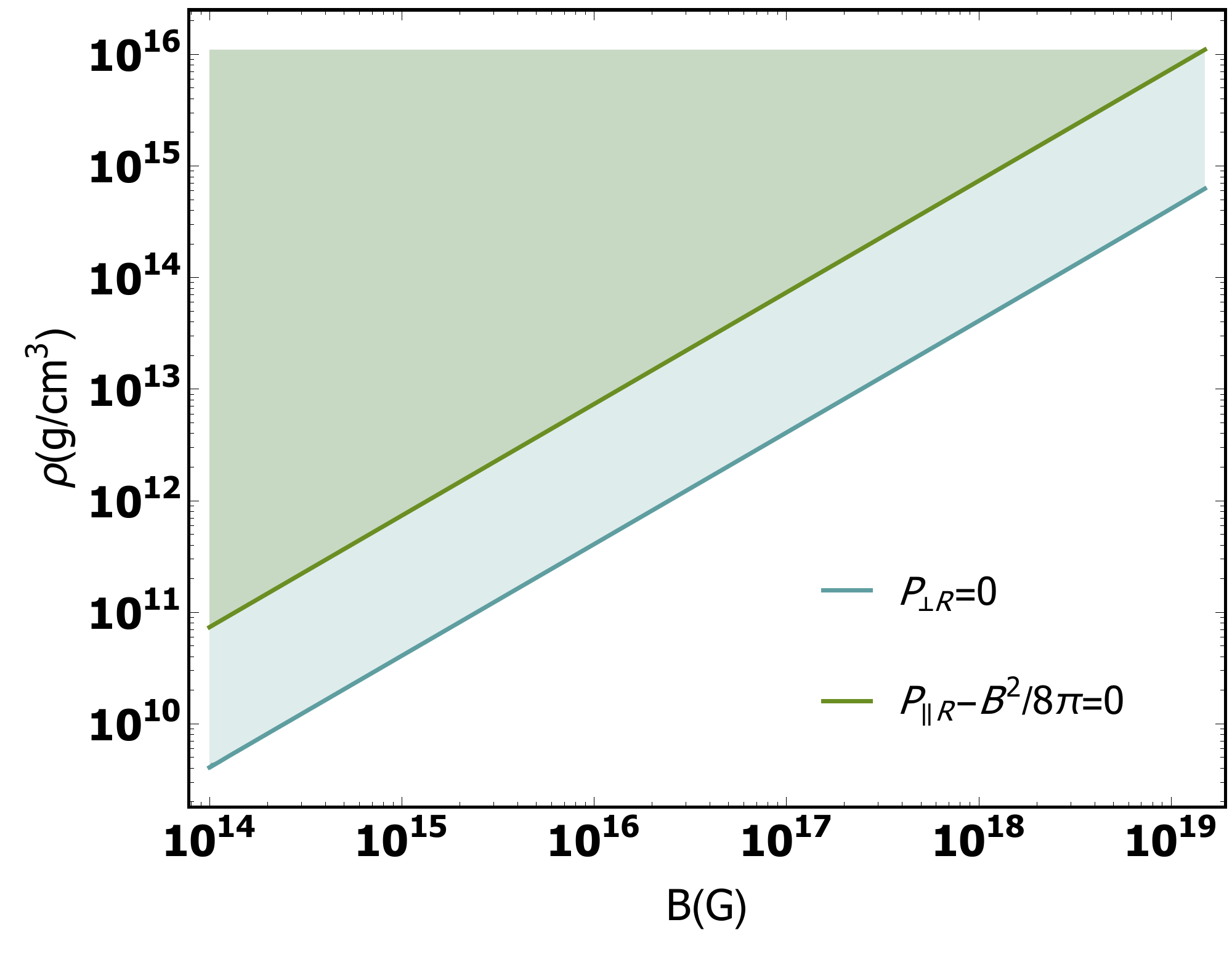}
	\caption{The parallel and perpendicular pressures as a function of the mass density for non relativistic, relativistic, and relativistic with Maxwell contribution EoS (left upper, right upper, and left lower panels respectively). Lower right panel: The phase diagram for the pressure instability in the boson mass density vs magnetic field plane. In the shadowed regions the gas is stable.}\label{EoSplotBcte}
\end{figure}

The effect of a constant magnetic field in the BEC star pressures is displayed in Fig.~\ref{EoSplotBcte} for $B=0$, $B=10^{17}$G and $B=10^{18}~G$. The upper panels correspond to the relativistic and the non-relativistic EoS without the Maxwell term, while lower left panel presents the relativistic EoS with Maxwell term. For the three EoS depicted, the high mass density region is dominated by boson-boson interactions and the deviations from the non-magnetic case are small. But as the density decreases, the magnetic field effects become more important.

For the non relativistic EoS (upper left panel of  Fig.~\ref{EoSplotBcte}), since the  parallel pressure is unaffected by the magnetic field, the curves for $P_{\parallel}^{nr}$ overlaps with the $B=0$ curve. On the contrary, in the region of lower densities, the non-relativistic perpendicular pressure separates from it as illustrate the solid curves. The mass density value at which the curves  $P_{\perp}^{nr}$ ends corresponds to the value at which $P_{\perp} ^{nr}= 0$ for a given $B$. For lower values of the boson mass density $P_{\perp}^{nr} < 0$ and the gas becomes unstable.

The relativistic perpendicular pressure diminishes with the mass density until becoming negative with the exact same dependence of the non relativistic case (upper right panel of Fig.~\ref{EoSplotBcte}).  This indicates that in what concerns $P_{\perp}$, the magnetic pressure $-{\mathcal M}B$ dominates over the vacuum one $-\Omega_{vac}$. Also, the similarities between $P_{\perp}^{nr}$ and $P_{\perp}^{r}$ indicates that ${\mathcal M}^{r} \cong	{\mathcal M}^{nr} $ for the magnetic fields here considered.

On the other hand, if the density is lowered below the value at which $P_{\perp}^{r} = 0$, the relativistic parallel pressure eventually separates from the $B=0$ curve and tends to the value $-\Omega_{vac}(b)$ which is constant with respect to the density (green dashed curve). This is the only difference between the non-relativistic and the relativistic cases, but since it happen for densities in the unstable pressure region (where $P_{\perp}<0$), we can conclude that  relativistic effects related to the magnetic field are not relevant in what concerns magnetized BEC stars. In fact, this could be previously noticed because the upper bound for magnetic field in NS interiors ($10^{18}$G) is two order less than the critical magnetic field $10^{20}$~G of the magnetized boson gas. In what follows, we use the relativistic EoS for calculations.

When Maxwell contribution is included in the EoS (left lower panel of Fig.~\ref{EoSplotBcte}) the role of pressures is exchanged. Now is the parallel pressure the one that becomes negative, while in the low density region, the perpendicular one diminishes until it reaches a constant value which is equal to the Maxwell contribution for the fixed value of $B$. In this case the pressure instability appears for higher values of the boson mass density and the difference between the pressures is greater than when Maxwell contribution is neglected.

In right lower panel of Fig.~\ref{EoSplotBcte}, the lines determining the onset of pressure instability $P_{\perp}^{r}(\rho,B)=0$ ($P_{\parallel}^{r}(\rho,B)-B^2/8\pi=0$) for the relativistic EoS without (with) Maxwell contribution are drawn, and the regions where the gas is stable has been shadowed. The pressure instability imposes a lower bound on the BEC star central densities, being the specific value of this bound an increasing function of $B$. Therefore, given a value of the magnetic field, only densities in the corresponding shadowed regions can sustain it inside the star. A remarkable feature of this plot is that, for the higher values of the magnetic fields, the limiting densities are in the order of nuclear saturation density or higher.

\subsubsection{EoS with self-generated magnetic field}
\label{sec4B}

Stellar magnetic fields are thought to decrease from the center to the surface of the star \cite{Lattimerprognosis,Chatterjee,ChatterjeeBprofiles}. In this section, and searching for a  more realistic description of the compact object, we take advantage of the fact that for low enough temperatures, spin one gases shows an spontaneous magnetization \cite{Yamada,Quintero2017IJMP,Lismary,Elizabeth}. This phenomenon is caused not by a coupling between the spin of the particles, but because bosons in the condensed phase are in the lowest energy state. For a magnetic gas this is a state with all the spins aligned. In consequence, low temperature spin one gases generate their own magnetic field $B_{sg}$ which is  proportional to the particle density \cite{Quintero2017PRC,Lismary}.
To compute the self-generated magnetic field $B_{sg}$ it is necessary to solve the equation

\begin{equation}\label{selfmag}
B_{sg} = 4 \pi {\mathcal M},
\end{equation}

\noindent the magnetization ${\mathcal M}$ is given by Eqs.(\ref{EoSNRtotal}) for the non relativistic case, and by Eqs.~(\ref{EoSRtotal}) in the relativistic one.

\begin{figure}[h!]
	\centering
	\includegraphics[width=0.42\linewidth]{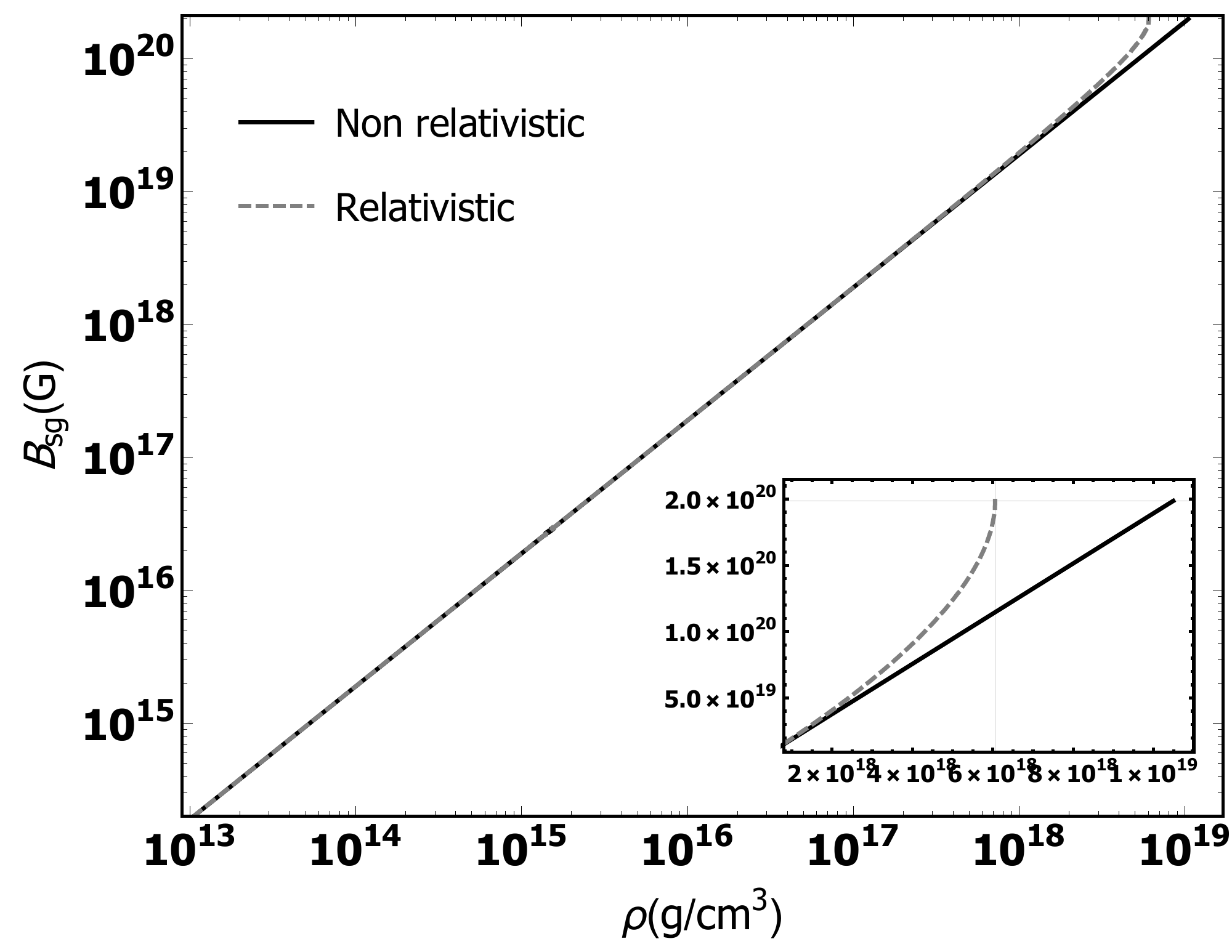}
	\includegraphics[width=0.42\linewidth]{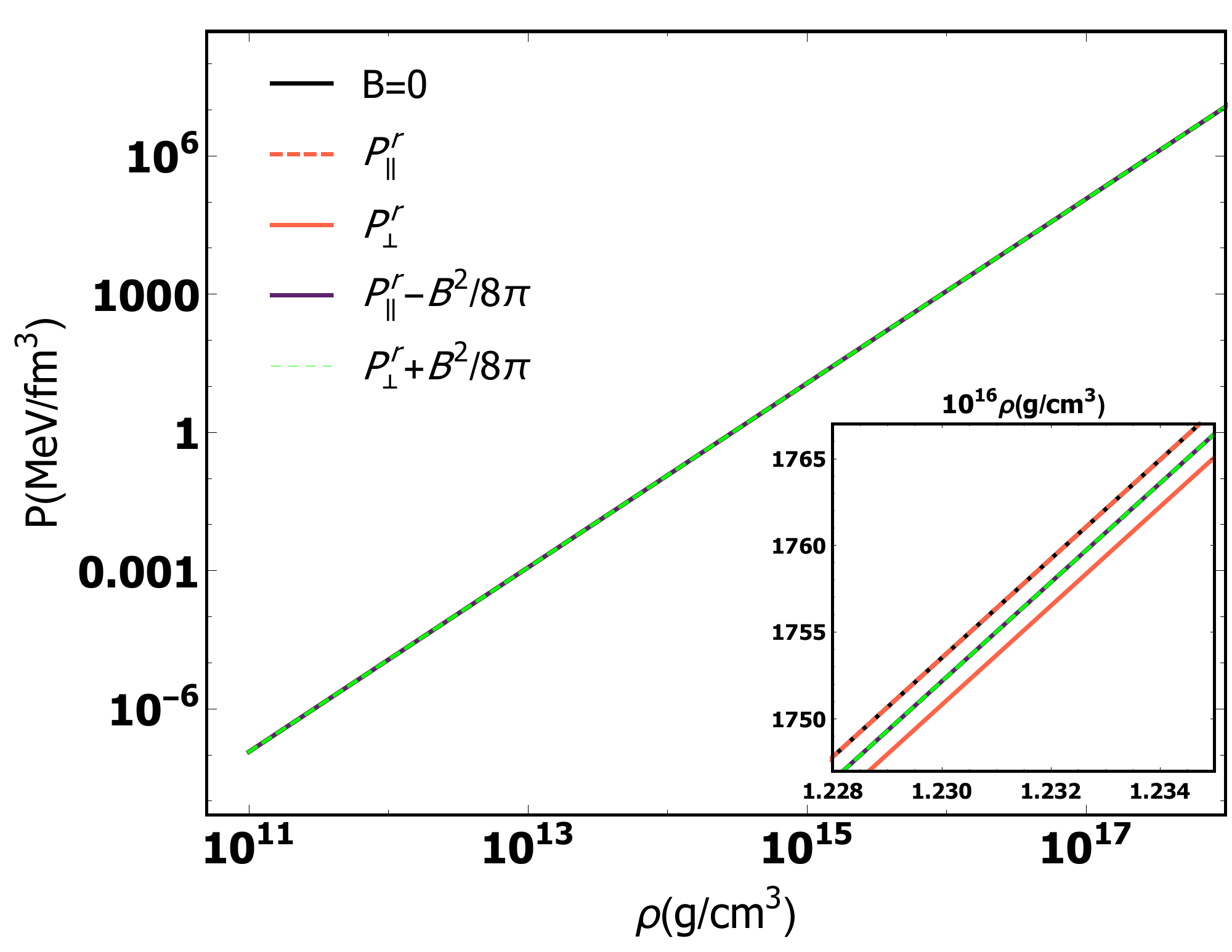}
	\caption{Left panel: the self-generated magnetic field as a function of the mass density for the relativistic and the non relativistic cases. Right panel: the parallel and perpendicular pressures as a function of the mass density for relativistic EoS with self-generated magnetic field.}\label{EoSplotBsg}
\end{figure}

For the non relativistic EoS, self magnetization equation becomes $B_{sg} = 4 \pi \kappa N$ and $B_{sg}$ increases linearly with the particle density, as shown in left panel of Fig.~\ref{EoSplotBsg}. Evaluating for some typical values of boson mass density, for instance $\rho =m N =10^{14}-10^{16}$g/cm$^{-3}$, one can see that the values of the self-generated field, $B=10^{15}-10^{17}$G, are in the order of those expected for neutron stars.

In the relativistic case, self-magnetization equation is $B_{sg} = 4 \pi \kappa/\sqrt{1-B_{sg}/B_c} N$ and $B_{sg}$ still increases with the mass density, (see the dashed curve in left panel of Fig.~\ref{EoSplotBsg}) but the dependence is not linear anymore. In addition, there exist a limiting particle density above which the self-magnetization condition is not fulfilled, and a maximum value of the magnetic field that can be reached through self-magnetization \cite{Quintero2017PRC}. For the bosons here considered, the maximum mass density and magnetic field are $\rho = 3.61 \times 10^{18}$g/cm$^3$ and $B=2/3 B_c = 1.98 \times 10^{20}$G respectively.
However, since maximum values of the inner magnetic field in NS are estimated to be in the order of $10^{18}$G, in the case of self-magnetization, the differences between the non relativistic and the relativistic cases are not relevant either.

In practice, to study the gas behavior under a self-generated magnetic field we only need to add Eq.~(\ref{selfmag}) to the EoS. The pressures with $B_{sg}$ are depicted in the right panel of Fig.~\ref{EoSplotBsg} as a function of the bosons mass density with and without taking into account the Maxwell contribution. A first significant feature in this plot is that the pressure instability disappear. This is a direct consequence of the decreasing of the magnetic field with the mass density. Also, note in the inset that having the Maxwell contribution erases the pressure anisotropy, although there is still a slightly difference with respect to the no magnetized EoS.

\section{Mass Radius relation for Magnetized and Self-magnetized BEC stars}

To have anisotropic EoS provokes a deformation  of the resulting compact object \cite{Paret2015}. In consequence, for an accurate description of magnetized stars it does not suffices to use the TOV structure equations, because they can only describe spherical stars. To consider the anisotropy properly, we will use the $\gamma$- structure equations, that describes axially symmetric stars provided they are spheroidal \cite{Samantha}. $\gamma$- structure equations are obtained starting from the metric

\begin{equation}\label{gammametric}
	ds^2 =  - \left[1-2M(r)/r\right]^{\gamma}dt^2 + \left[1-2M(r)/r\right]^{-\gamma}dr^2 + r^2\sin^2\theta d\phi^2 + r^{2}d\theta^2,
\end{equation}
that describes an object with axial symmetry in spherical coordinates \cite{Herrera:1998eq, Zubairi:2017yna, Zubairi:2017gvp}. In Eq.~(\ref{gammametric}), $\gamma = z/r$ parametrizes the polar radius $z$ in terms of the equatorial one $r$ and  accounts for the axial deformation of the object.

With the use of Eq.~(\ref{gammametric}) and computing the mass of the star as for an spheroid, the following structure equations are obtained \cite{Samantha}

\begin{subequations}\label{gTOV}
	\begin{eqnarray}
	&& \frac{dM}{dr}=\gamma r^{2}(E_{\perp}+E_{\parallel})/2, \label{gTOV1}\\
	&& \frac{dP_{\parallel}}{dr}=-\frac{(E_{\parallel}+P_{\parallel})[\frac{r}{2}+r^{3}P_{\parallel}-\frac{r}{2}(1-\frac{2M}{r})^{\gamma}]}{r^{2}(1-\frac{2M}{r})^{\gamma}}, \label{gTOV2}\\
	&& \frac{dP_{\perp}}{dz}=\frac{1}{\gamma}\frac{dP_{\perp}}{dr} =-\frac{(E_{\perp}+P_{\perp})[\frac{r}{2}+r^{3}P_{\perp}-\frac{r}{2}(1-\frac{2M}{r})^{\gamma}]}{\gamma r^{2}(1-\frac{2M}{r})^{\gamma}}, \label{gTOV3}
	\end{eqnarray}
\end{subequations}

\noindent where $M(r)$ is the total mass enclosed in the spheroid of equatorial radius $r$ and, at each integration step, $E_{\parallel}$ and $E_{\perp}$ are computed using the parametric dependence of the energy in each pressure derived from Eqs.(\ref{EoSRtotal}). Eqs.~(\ref{gTOV}), are subjected to the initial conditions $E_0 = E(r=0)$, $P_{\parallel_0} = P_\parallel(r=0)$, and $P_{\perp_0} = P_\perp(r=0)$ where $E_0$, and $P_{\perp_0}$ and $P_{\parallel_0}$ are taken form the EoS, while the condition $P(R) = 0$ defines the star equatorial radius from which the polar one $Z = \gamma R$ and the total mass $M(R)$ are computed.

To obtain Eqs.~(\ref{gTOV}), the dependence on the angular variables was canceled and the anisotropy enters on them only through $\gamma$ which is suppose to be near $1$. In a previous work, \cite{Samantha}, have been proposed to interpret $\gamma$ as the ratio between the parallel and perpendicular central pressures, $P_{\parallel_0}$ and $P_{\perp_0}$ respectively

\begin{equation}
\gamma=\frac{P_{\parallel_0}}{P_{\perp_0}}, \label{gamma}
\end{equation}

\noindent as a manner to connect the geometry and the physical properties of the system. Eq.~(\ref{gamma}) means that the shape of the star is determined by the anisotropy in its center and yields reasonable results, at least as a first approximation, as was illustrated for magnetized white dwarfs \cite{Samantha}.

\subsection{Masses and radii with constant magnetic field: deformation}

\begin{figure}[h!]
	\centering
	\includegraphics[width=0.42\linewidth]{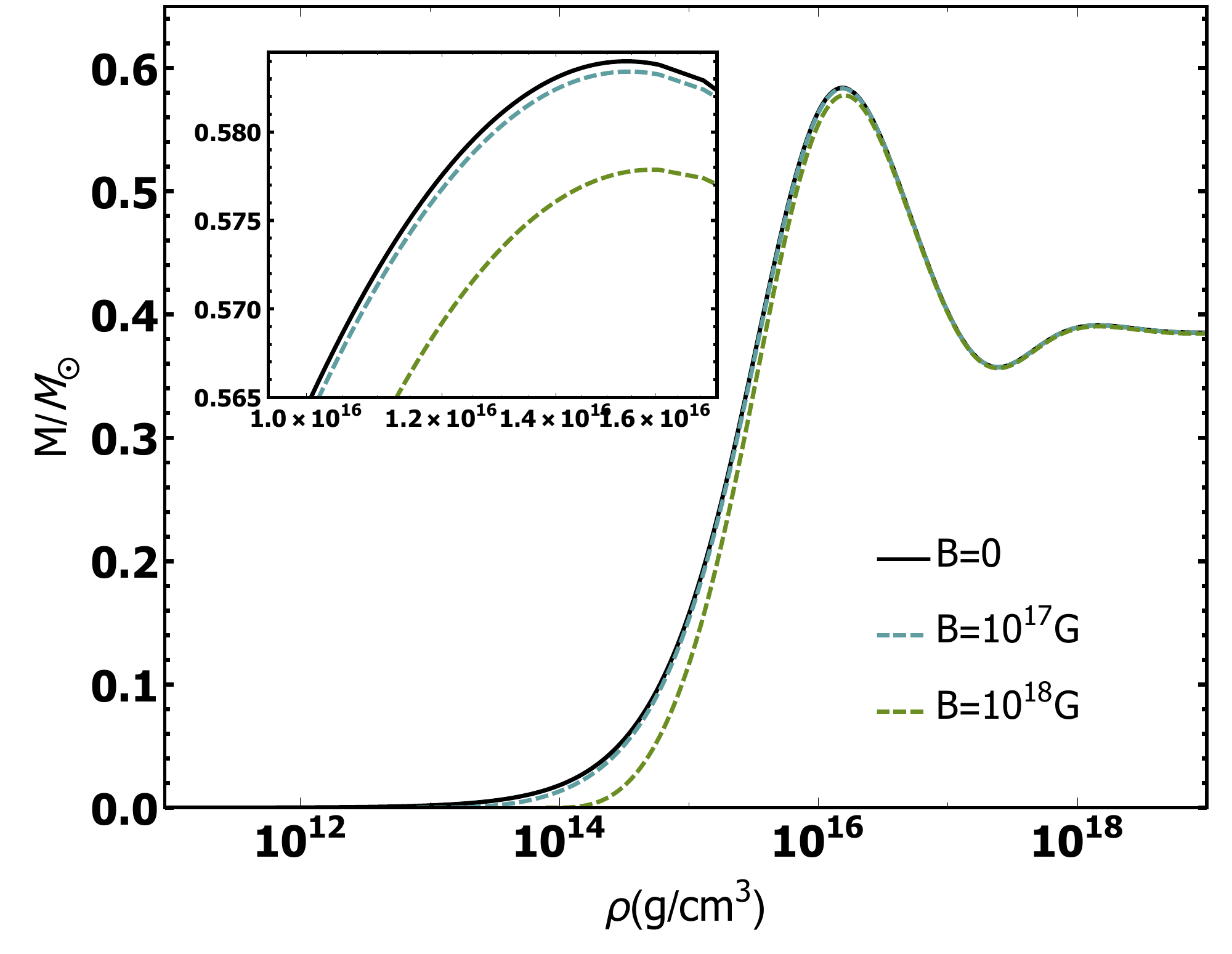}
	\includegraphics[width=0.42\linewidth]{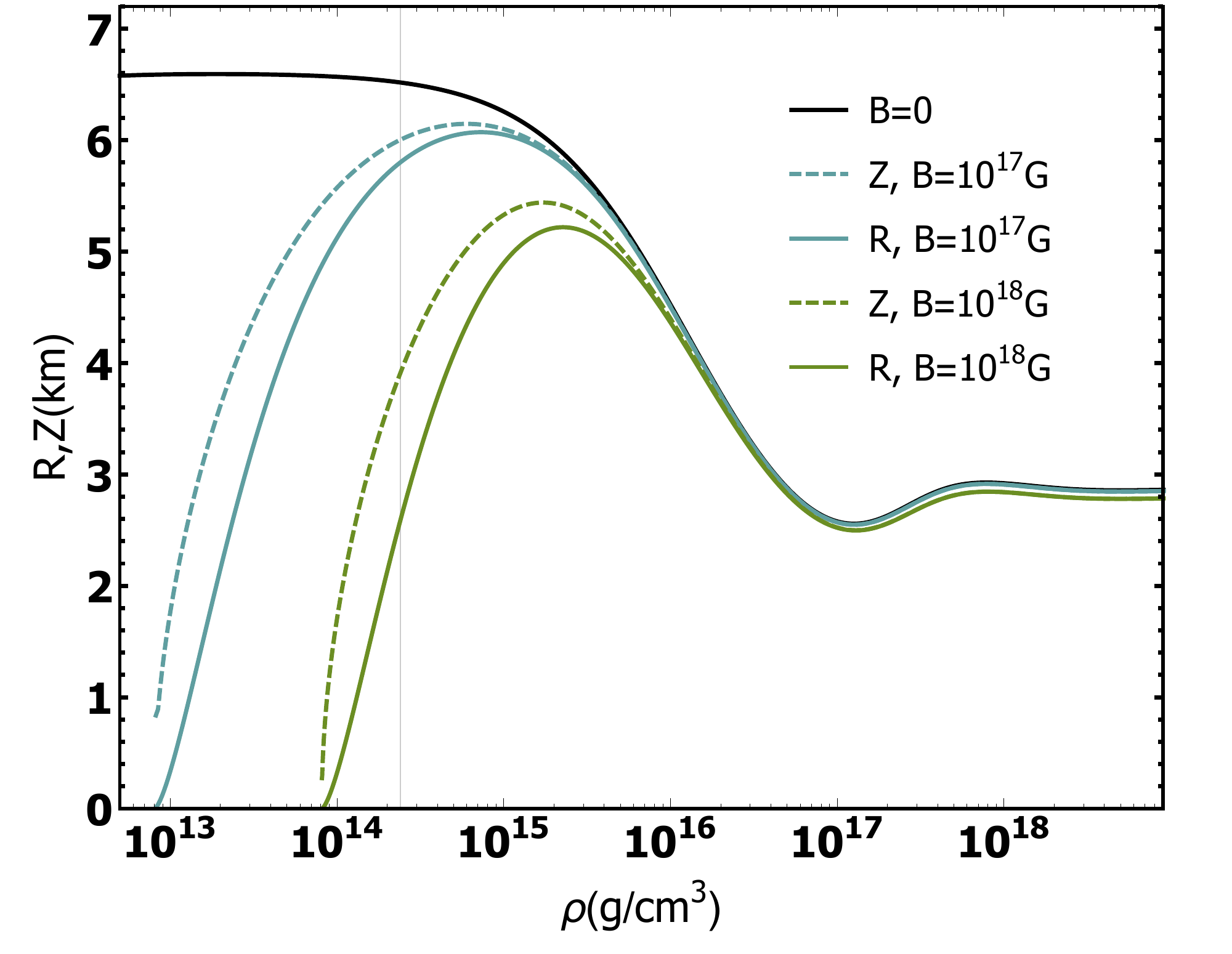}\\
	\includegraphics[width=0.42\linewidth]{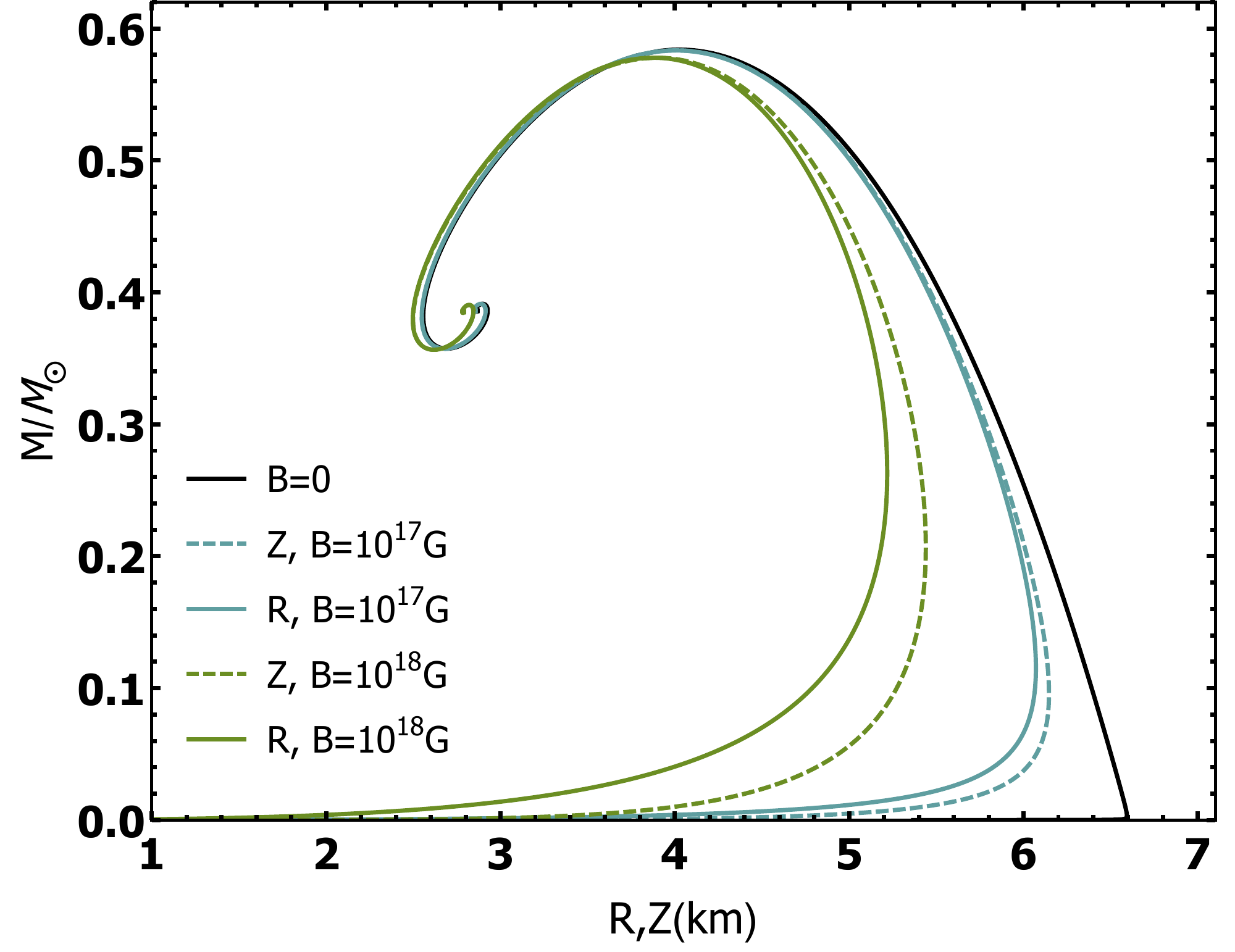}
	\includegraphics[width=0.42\linewidth]{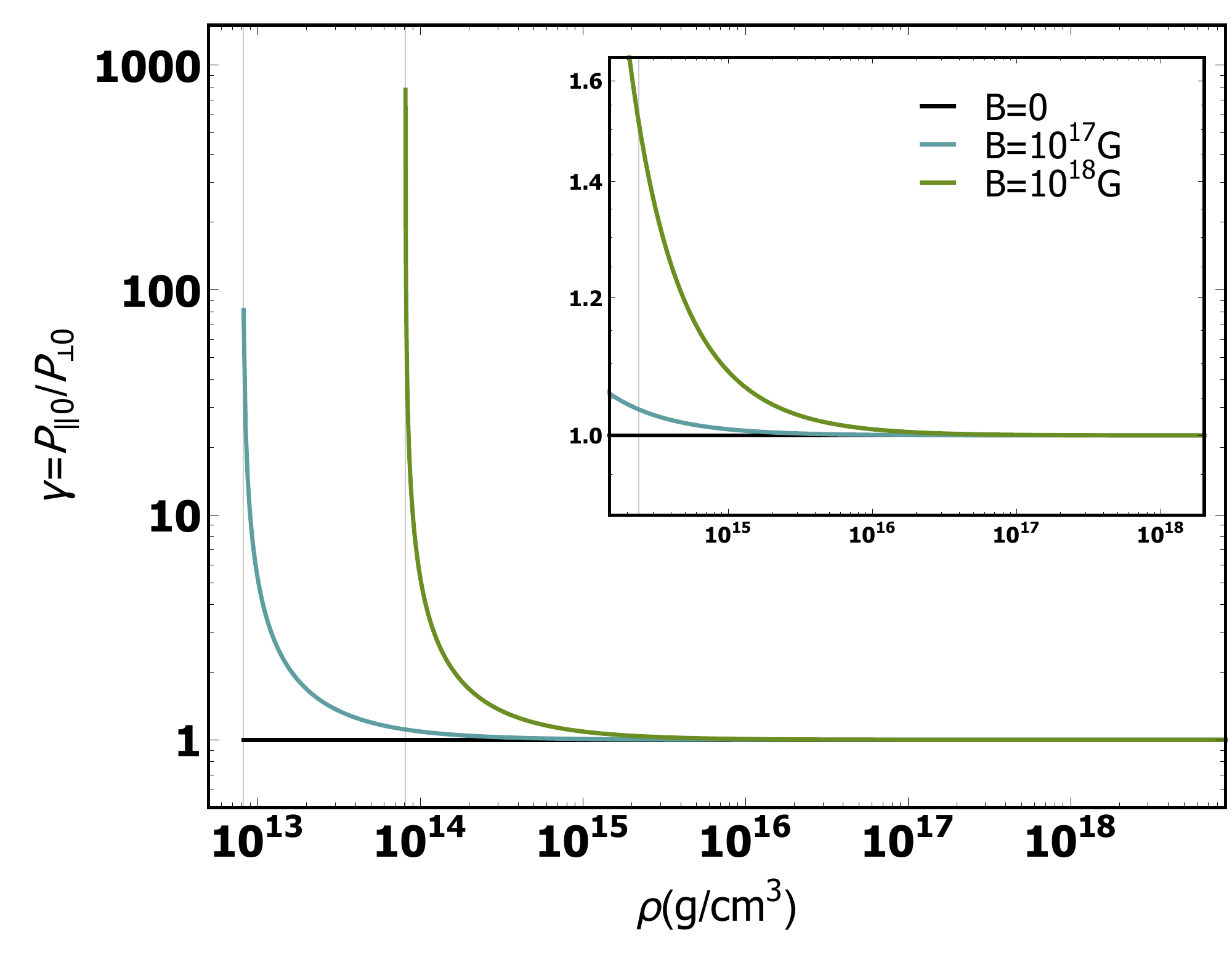}	
	\caption{The results of solving the $\gamma$-structure equations for the EoS without Maxwell contribution. Upper  panels: the total mass and the equatorial and polar radii  of the star as a function of the central mass density. Vertical line pinpoints $\rho_{nuc}$. Lower left panel: mass-radii relations with the equatorial (solid lines) and the polar (dashed lines) radius. Lower right panel: the parameter $\gamma$ as a function of the central mass density. The verticals lines signal the densities at which $P_{\parallel} =0$ and $\gamma \rightarrow \infty$ for a given magnetic field. The vertical line in the inset signals $\rho_{nuc}$.}\label{mrzro}
\end{figure}

In this section we discuss the results of solving the structure equations Eqs.~(\ref{gTOV}) for the EoS Eqs.~(\ref{EoSRtotal}) with constant magnetic  field. Fig.~\ref{mrzro} shows them for $B=0$, $B=10^{17}$G and $B=10^{18}$G without Maxwell contribution. As can be noticed in the upper left panel, in the $B=0$ case, the $M$ vs $\rho_0$ curve shows a region of stable configurations of BEC stars whose maximum mass is $M \approx 0.58 M_{\odot}$, at a radius of $R \approx 4.2$km and a central boson mass density $\rho_0 \approx 1.5 \times 10^{16}$g/cm$^3$ This result is in agreement with the one obtained in \cite{Latifah}. A constant magnetic field does not changes the form of the $M$ vs $\rho_0$ curve, but slightly decreases the mass. The decreasing of the mass is bigger at low densities, although it is also appreciable at the highest ones, as can be appreciated in the inset plot in the upper left panel of Fig.~\ref{mrzro}.

The influence of the magnetic field on the size and shape of the BEC stars is more dramatic. It not only deforms the object, but also diminish its size. Both effects can be seen in upper right and lower left panels of Fig.~\ref{mrzro}. Since $\gamma =z/r=P_{\parallel_0}/P_{\perp_0} $, the polar radius $Z$ is always bigger than the equatorial one $R$ (because when Maxwell contribution is ignored $P_{\parallel}>P_{\perp}$). This means that the resulting star is a prolate object. In the low density region there is an enormous deviation from the $B=0$ curve, with the radii decreasing with the density instead of tending to a  constant value. This decrement is directly related with the instability region of pressures showed by the EoS, in a way that, as the central perpendicular pressure approaches to zero, the stars become smaller and less massive.

Let us note that as the central mass density at which $P_{\perp_0}=0$ is reached, $\gamma \rightarrow \infty$ departing from the assumption $\gamma \cong 1$. Nevertheless, since the densities at which $\gamma$ diverges are under the nuclear saturation density (see lower right panel of Fig.~\ref{mrzro}), they are not expected to exist inside our stars\footnote{Although we are showing the results of integrating the structure equations for all the region of pressure stability, since the NS core is defined by $\rho>\rho_{nuc} \cong 2.4 \times 10^{14}$g/cm$^3$, the solutions for $\rho_0 < \rho_{nuc}$ must be ruled out.}.
At $\rho_{nuc}$ the values of $\gamma$ are quite moderated, as shown in the inset of lower right panel of Fig.~\ref{mrzro}, but deformation is still visible. In the high densities region, $\gamma \rightarrow 1$ and the stars are almost spherical.

\begin{figure}[h!]
	\centering
	\includegraphics[width=0.42\linewidth]{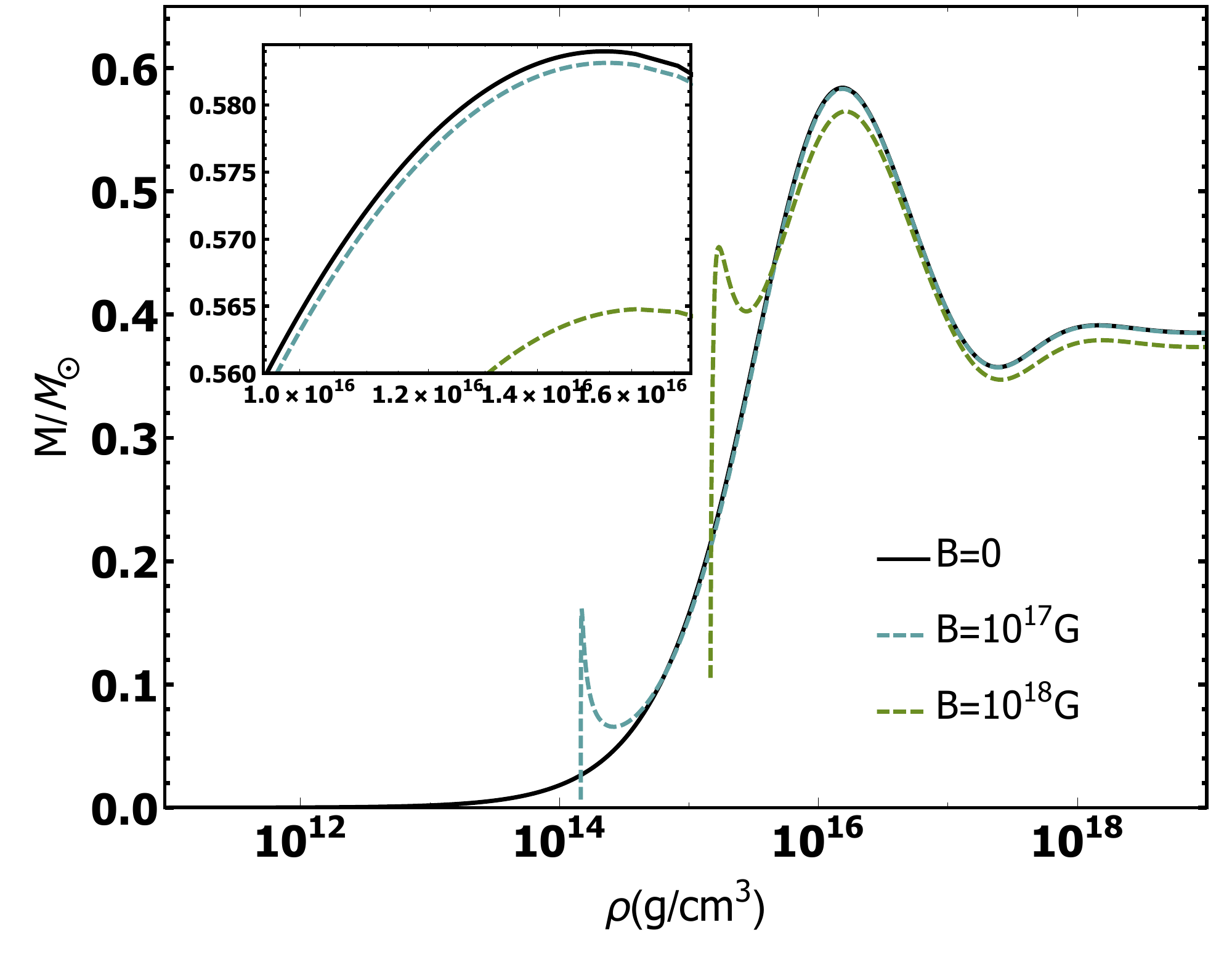}
	\includegraphics[width=0.42\linewidth]{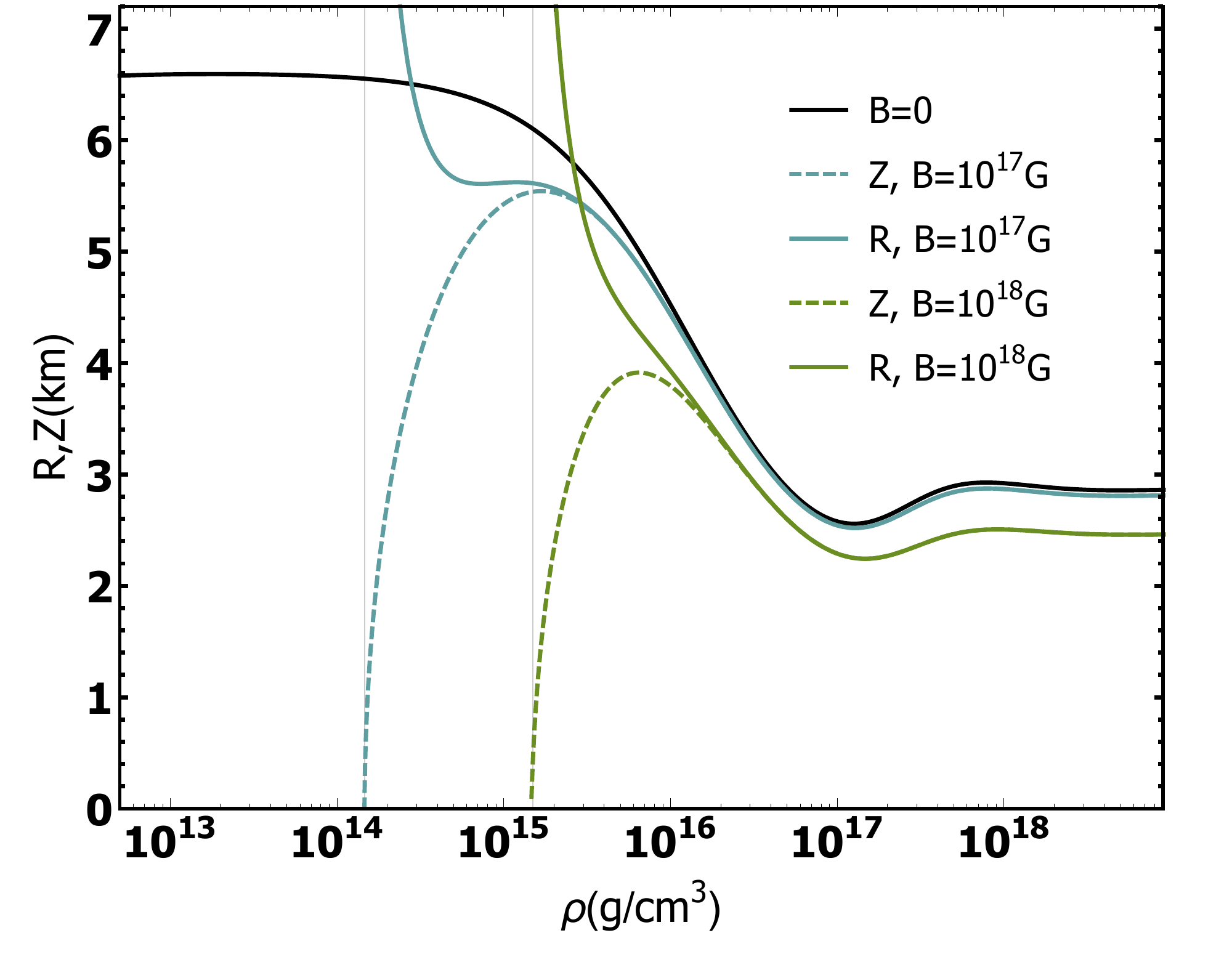}\\
	\includegraphics[width=0.42\linewidth]{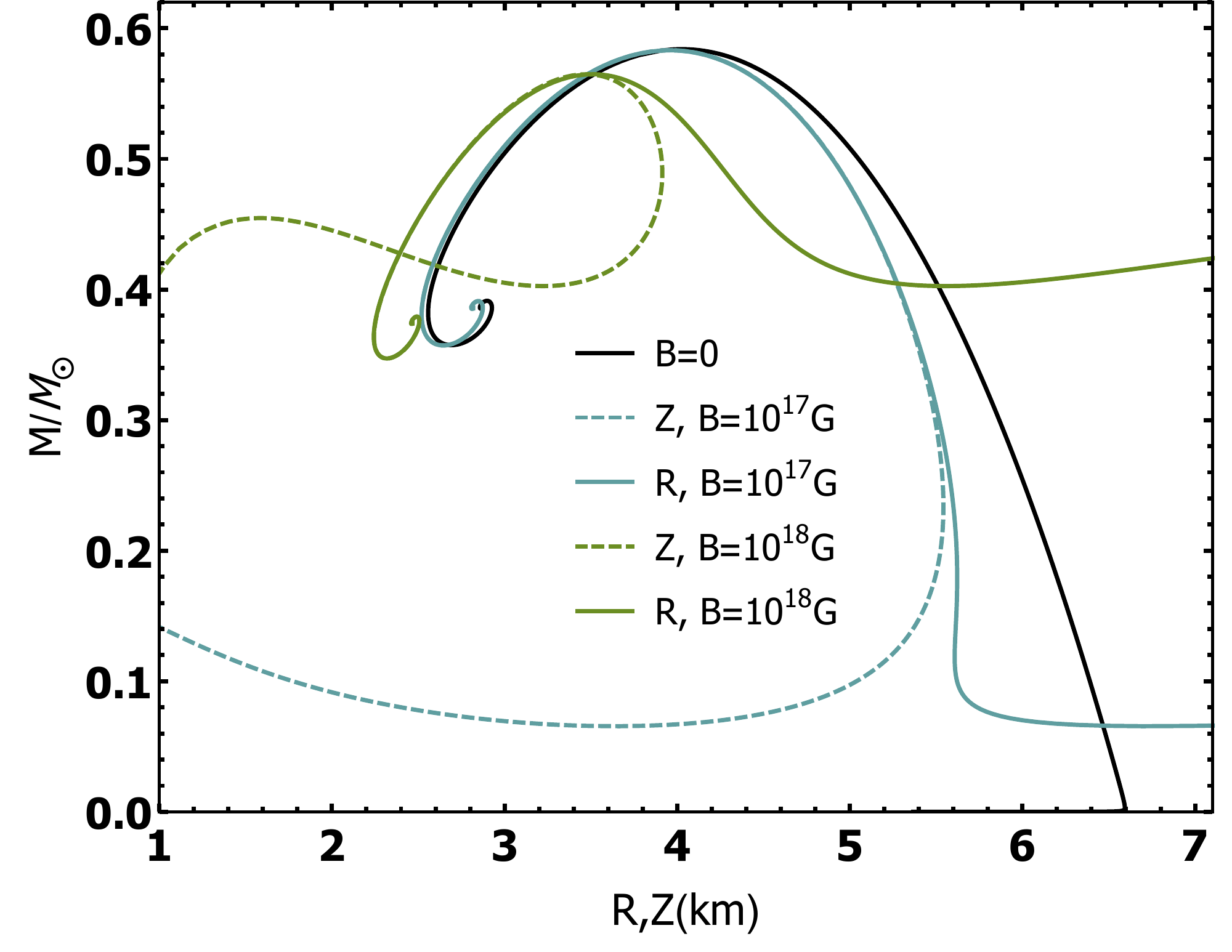}
	\includegraphics[width=0.42\linewidth]{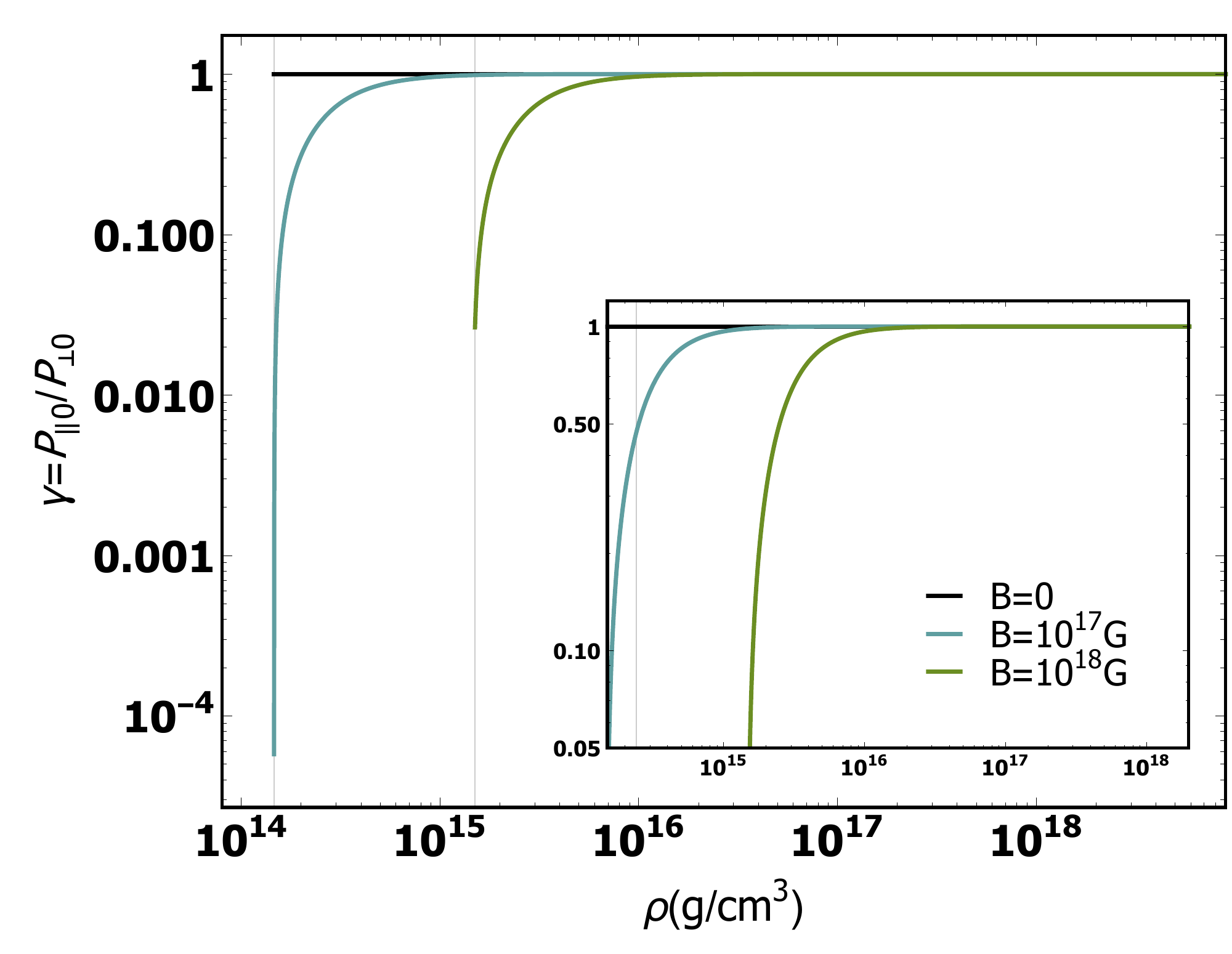}	
	\caption{The results of solving the $\gamma$-structure equations for the EoS with Maxwell contribution. Upper  panels: the total mass and the equatorial and polar radii  of the star as a function of the central mass density. Vertical line pinpoints $\rho_{nuc}$. Lower left panel: mass-radii relations with the equatorial (solid lines) and the polar (dashed lines) radius. Lower right panel: the parameter $\gamma$ as a function of the central mass density. The verticals lines signal the densities at which $P_{\parallel} =0$ and $\gamma \rightarrow 0$. The vertical line in the inset signals $\rho_{nuc}$. }\label{mrzroBcuadrado}
\end{figure}

The stable mass-radius configurations obtained with Maxwell contribution are shown in Fig.~\ref{mrzroBcuadrado}. As we saw for the EoS, Maxwell contribution reinforces the effects of the magnetic field, so now the decreasing of the mass and size of the BEC stars is higher and easily noticeable for almost all densities. In this case $P_{\parallel}<P_{\perp}$, therefore $\gamma<1$, the equatorial radius is always bigger than the polar one and the star is an oblate object. The equatorial radius increases with decreasing density, while the polar radius diminishes, their difference increases up to four orders for $B=10^{17}$~G (see lower right panel of Fig.~\ref{mrzroBcuadrado}). This behavior, as well as the new peak in the mass-central mass density curve, is related to the fact that when $P_{\parallel_0} \rightarrow 0$, $\gamma \rightarrow 0$ as shows the lower right panel of Fig.~\ref{mrzroBcuadrado}. The limit $\gamma \rightarrow 0$ is outside the range of validity of the structure equations, and in fact, setting $\gamma = 0$ in Eq.~(\ref{gammametric}) transforms the $\gamma$ metric into the flat Minkowski space-time \cite{MalafarinaHerrera}, therefore solutions of the structure equations around this limit does not have physical interest.

\subsection{Self-magnetized BEC stars and inner magnetic field profiles}

\begin{figure}[h]
	\centering
	\includegraphics[width=0.42\linewidth]{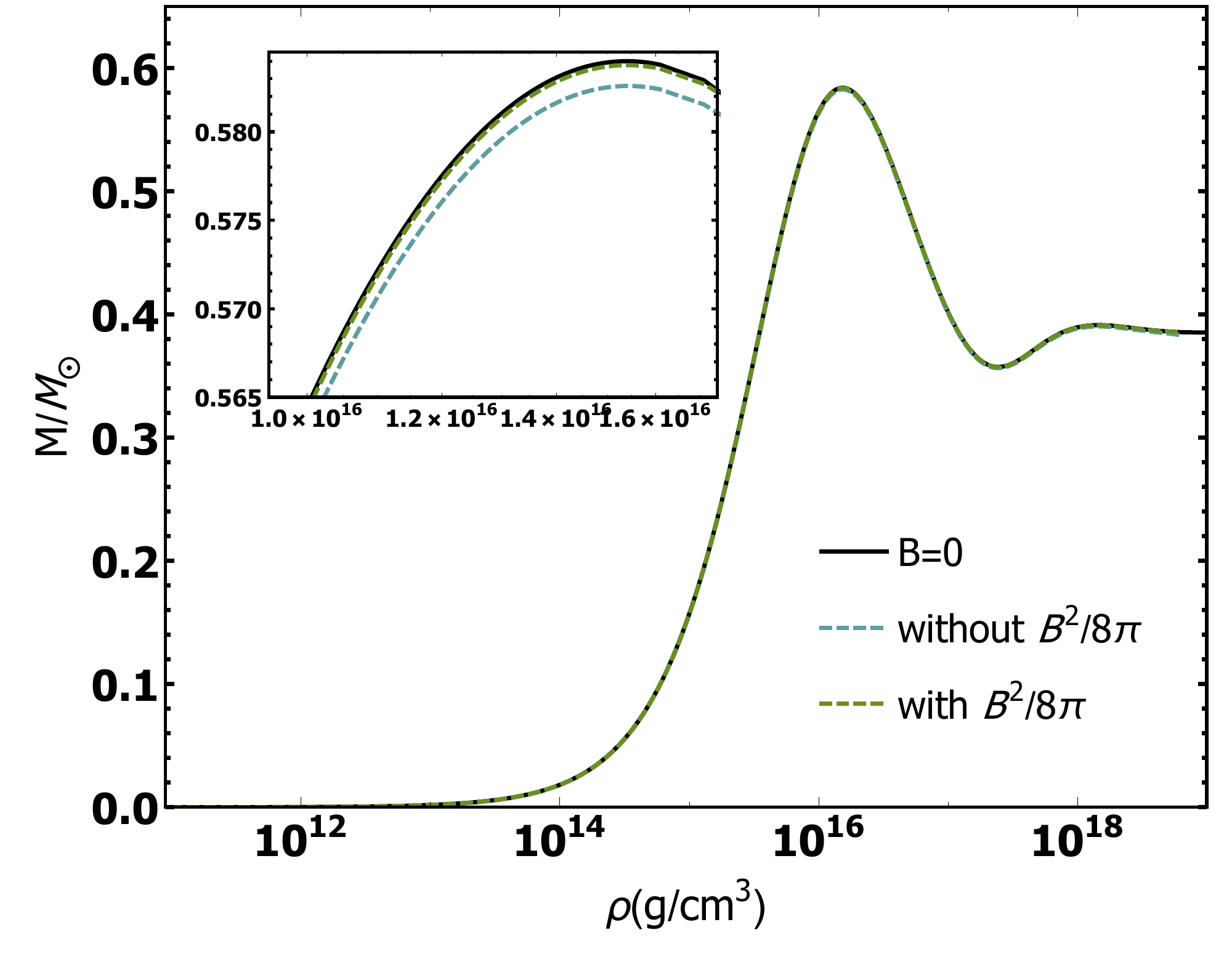}
	\includegraphics[width=0.42\linewidth]{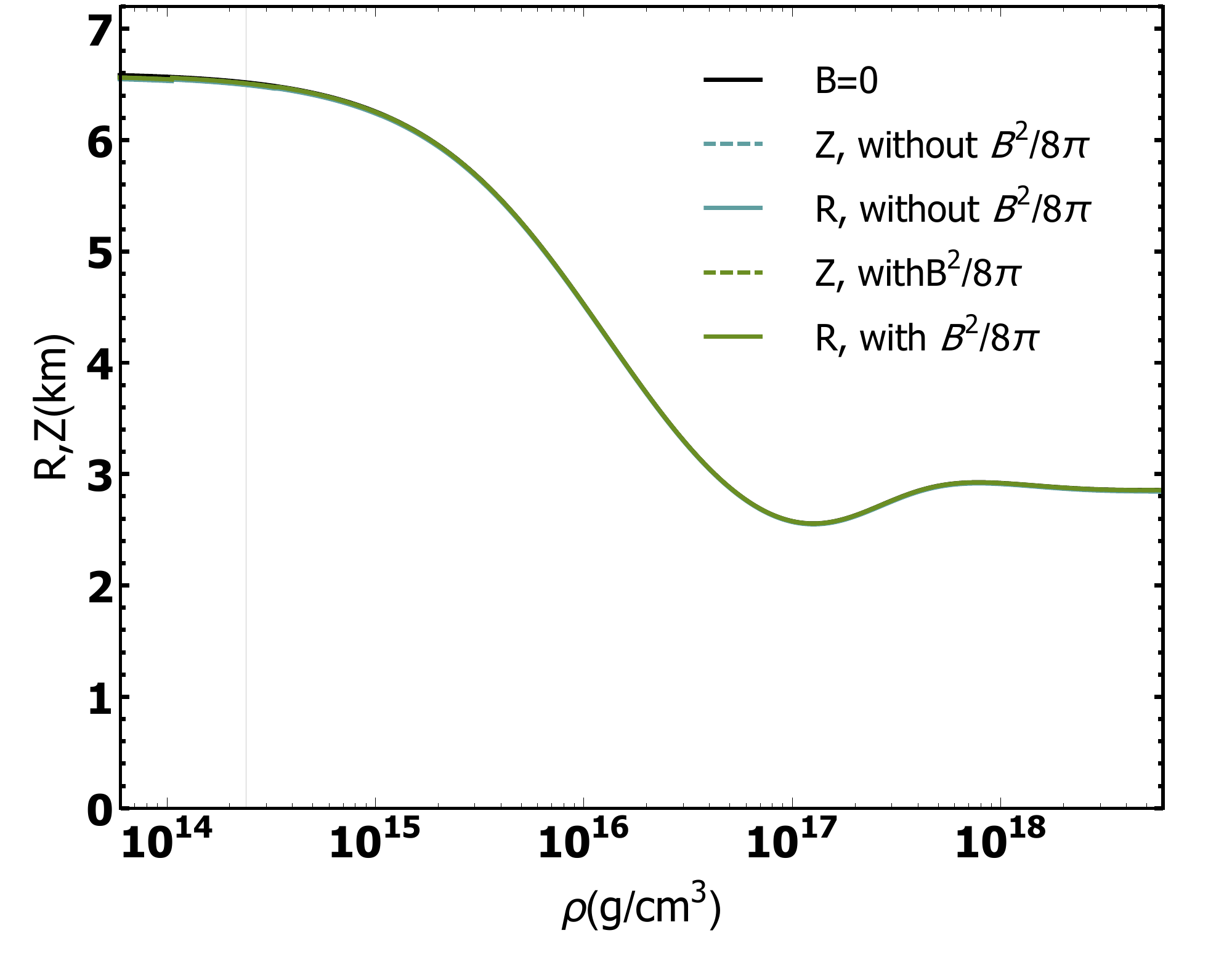}\\
	\includegraphics[width=0.42\linewidth]{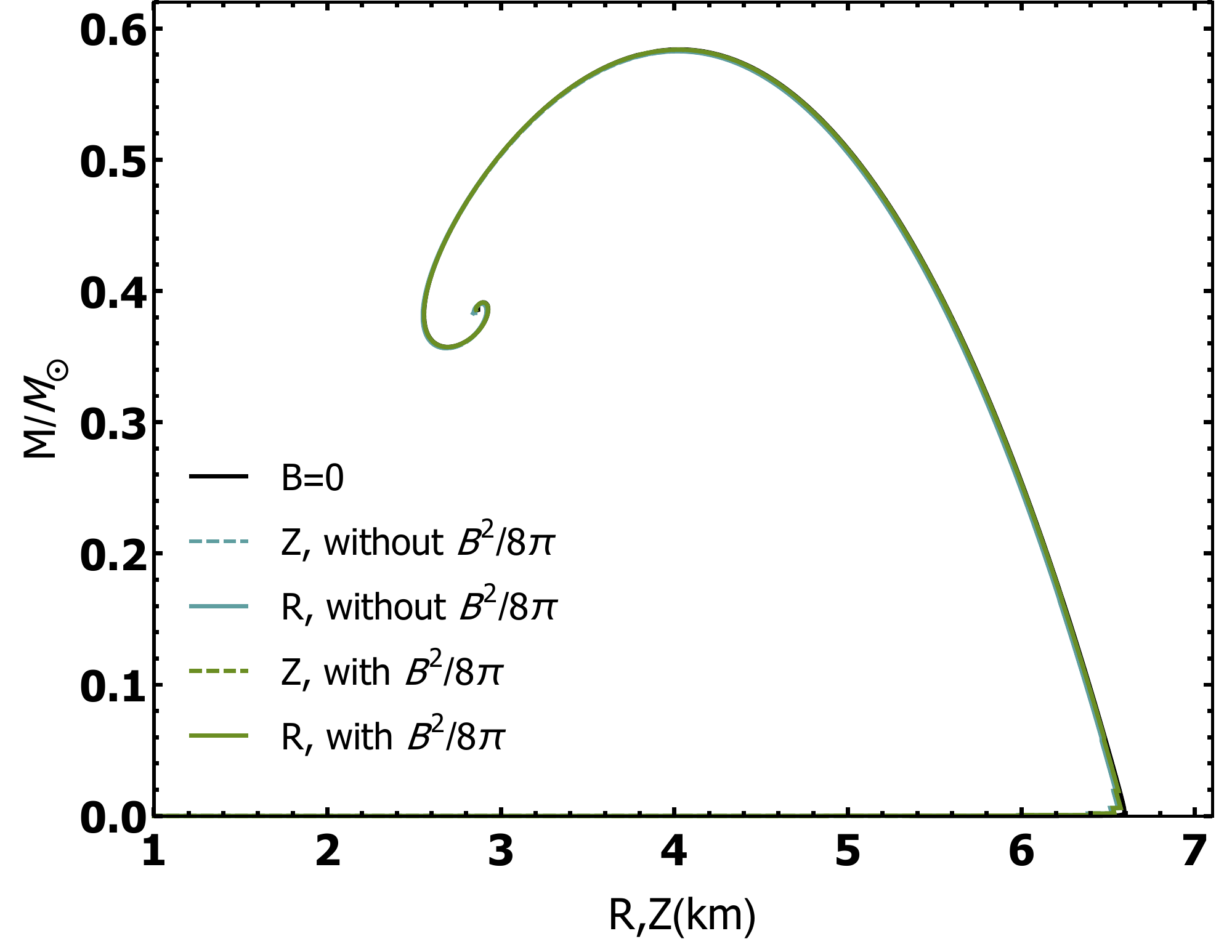}
	\includegraphics[width=0.42\linewidth]{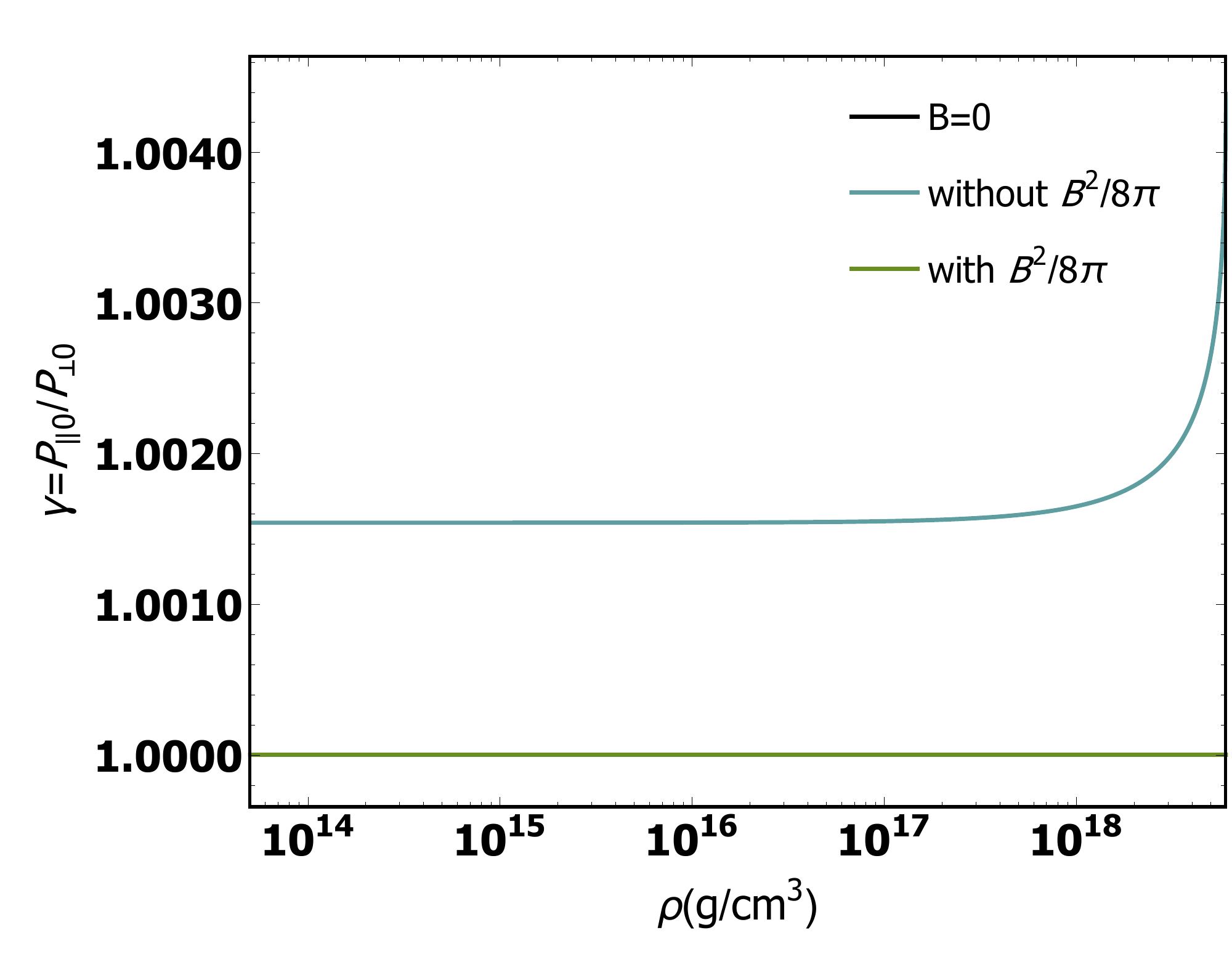}	
	\caption{The results of solving the $\gamma$-structure equations for the EoS with self-generated magnetic field. Upper panels: the total mass and the equatorial and and polar radii  of the star as a function of the central mass density.  Vertical line signals $\rho_{nuc}$. Lower left panel: mass-radii relations with the equatorial (solid lines) and the polar (dashed lines) radius. Lower right panel: the parameter $\gamma$ as a function of the central mass density.}\label{mrzroBsg}
\end{figure}

In this subsection we solve Eqs.~(\ref{gTOV}) for the EoS Eqs.~(\ref{EoSRtotal}) with the magnetic field given as a function of the mass density through Eq.~(\ref{selfmag}). The results are shown in Fig.~\ref{mrzroBsg}. The curves corresponding to the self-magnetized BEC stars almost perfectly overlaps with the $B=0$ ones, and the influence of magnetic field on the mass, size and shape of the star is really small (see the inset on upper left panel of Fig.~\ref{mrzroBsg}). The reason for that comes from the dependence of the self-generated magnetic field on the boson mass density, that diminish the anisotropy in the EoS in a way that the instability region of the pressures never appears. As a consequence, the resulting self-magnetized BEC stars deviates slightly from the spherical shape and their masses are barely diminished, being $\gamma \cong 1$ for all central mass densities.

\begin{figure}[h]
	\centering
	\includegraphics[width=0.42\linewidth]{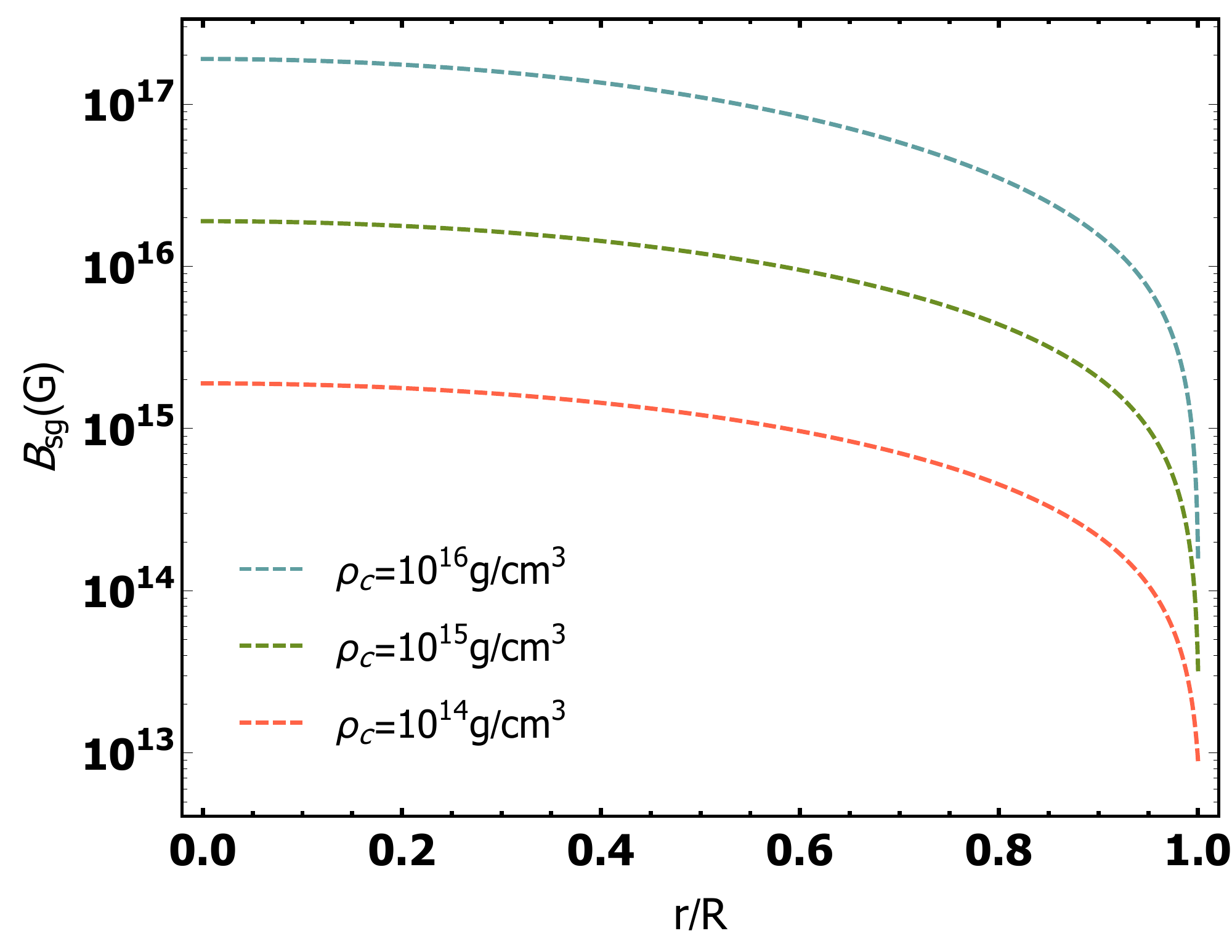}
	\includegraphics[width=0.42\linewidth]{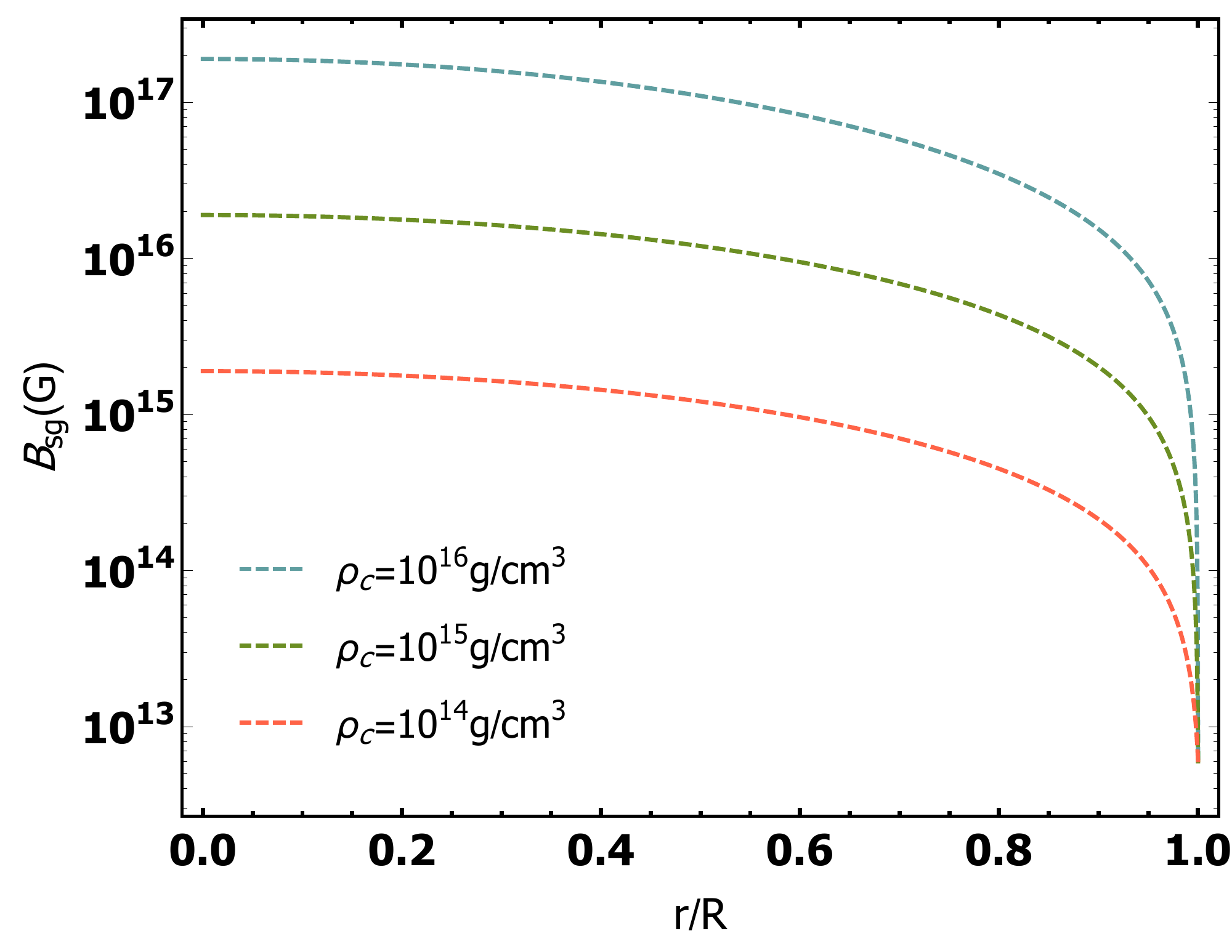}
	\caption{The magnetic field in the interior of the BEC star as a function of the equatorial radius for several values of the star's central mass density. Left panel: EoS without Maxwell contribution; right panel: EoS with Maxwell contribution }\label{Bprofiles}
\end{figure}

The use of Eq.~(\ref{selfmag}) along with the EoS, allows to compute the magnetic field intensity self-consistently during the integration of the structure equations.
The magnetic field profiles are depicted in Fig.~\ref{Bprofiles} as a function of the equatorial radius for various central mass densities. Left (right) panel of the figure shows the curves for the case without (with) the inclusion of Maxwell contribution. At the center of the star the values of the magnetic field are the same provided we have the same central mass density. But the decrease of the magnetic field at the star surface is bigger when Maxwell contribution is included, the variation being around three (four) orders for the EoS with (without) Maxwell term (see Table~\ref{T1}). In both cases the values at center as well as the ones at star surface are in the orders of those estimated for NS \cite{Malheiro:2013loa,1991ApJ...383..745L,Lattimerprognosis}.

\begin{table}[ht]
\begin{tabular}{|c|c|c|r|}\toprule
 $\rho _0 (\text{g}/\text{cm}^3)$ & $B_0\text{(G)}$&$B_s\text{(G)}$ & Maxwell \\
 \hline
 &                 & $5.98856\times10^{12} $ & with    \\
 \cline{3-4} $10^{14}$                       & $1.89376\times 10^{15}$ & $8.86556\times 10^{12}$    & without \\
 \hline
 &                 &$5.98856\times 10^{12}$ & with\\
 \cline{3-4} $10^{15}$                      & $1.89381\times 10^{16}$ & $3.27446\times 10^{13}$     & without \\
		\hline
 &                 &$5.98855\times 10^{12}$ & with    \\
 \cline{3-4} $10^{16}$                      & $1.89435\times 10^{17}$&  $1.62406\times 10^{14}$     & without \\
 \toprule
\end{tabular}
\caption{The magnetic field intensity at the center and the surface of self-magnetized BEC stars for different values of the central mass density.}
\label{T1}
\end{table}

The results of this section validates the spin one bosons as a possible candidates for magnetic field source not only in the case of magnetized BEC stars, but also for other models of NS where certain amount of them are present. One of the merits of our scheme is, indeed, the fact that the vector boson gas can give rise to a self-generated magnetic field consistent with astronomical observations. In addition, the magnetic field profiles obtained within this framework stem naturally from the solution of the structure equations. By fixing its orientation, the self-magnetized BEC star magnetic field is thus a first principle quantity, free from any heuristic assumptions (e.g., at difference with Ref.~\cite{ChatterjeeBprofiles} approach).

\section{Conclusions}

We have obtained the EoS and the mass-radius relation of self-magnetized and magnetized BEC stars. Such stars are formed by a magnetized gas of interacting spin one bosons composed by the spin parallel pairing of two neutrons in the interior of neutron stars, so we are assuming in our model that, at least at some stage of their evolution, neutron stars cores might be composed by matter in this form.

The thermodynamic description of the gas was done for relativistic and non relativistic bosons. However, since we worked for $B\ll B_c$, the differences between these regimes were insignificant. To include the magnetic field we assumed that boson-boson and boson-magnetic field interactions are independent, and considered two configurations: in one the magnetic field is taken as constant in the interior of the star and externally fixed, while in the other it is produce by the bosons by self-magnetization and depends on their mass density. The consequences of including or not the Maxwell contribution to the energy density and the pressures are also studied.

The main effect of the magnetic field in the BEC star  EoS is the splitting of the pressures in two components, one along and the other perpendicular to the magnetic axis. To obtain the structure of the magnetized BEC stars without neglecting the anisotropy in the EoS, we have used  the axially symmetric structure equations presented previously in \cite{Samantha}. They allowed us to obtain the mass and radius of stable spheroidal objects whose deformation, given by the parameter $\gamma$, is proportional to the ratio of the star central pressures. In spite of the approximate character of these structure equations, they give reasonable results as a first approximation indicating whether the deformation is important or not.

For constant magnetic field, the anisotropy in the pressures is significant, and increases when Maxwell contribution is included. Furthermore, in the low density region, the smallest pressure becomes negative and the gas unstable. So, taking the magnetic field as constant imposed a lower bound in the central mass densities that are needed to sustain it inside the star. The limiting density is a increasing function of the magnetic field, and although we make the distinction of high/low densities, its values are around or above $\rho_{nuc}$.

The magnetized (non-spherical) BEC stars are smaller than their non-magnetic (spherical) analogues. The star deformation is more pronounce for low densities. The oblateness or prolateness of the resulting compact object depends on which direction the largest pressure is exerted. In our model, this depends on whether the Maxwell contribution is or not taken into account. The extreme behavior of masses, radii and shape observed in the proximity of the pressure instability regions are for sure consequence of the EoS, but are also related to the the properties of $\gamma$-structure equations in the limits $\gamma \rightarrow \infty$ and $\gamma \rightarrow 0$. Therefore, further studies are needed to gain a deeper insight on this question.

When the magnetic field is created by self-magnetization, the anisotropy in the EoS is insignificant and the pressure instability region never appears. As a consequence, a self-generated magnetic field produces small changes and almost no deformation in the compact object structure. Nevertheless, taking the magnetic field as produce by the bosons allows to compute its intensity in the interior of the star during the integration of the structure equations. The obtained magnetic field profiles, as well as their extreme values, are in agreement with the observations and theoretical predictions for neutron stars. This support the idea that spin one bosons might work as magnetic field sources for compact objects. We would like to remark that self-magnetization provides a natural way of introducing a magnetic field that is directly produced by the matter that composes the star.

\section{Acknowledgements}

The authors thank Pedro Bargueño for interesting discussions about some aspects of this work and Gabriel Gil for useful comments on the manuscript. The authors have been supported by the grant of PNCB-MES Cuba No. 500.03401. G.Q.A. expresses gratitude to the Service de Coopération et d'Action Culturelle (SCAC) of the Embassy of France in Cuba, to LUTH at the Paris Observatory and to the Abdus Salam ICTP, for the support and hospitality during the final stage of this work. D.M.P. has been also supported by a research grant of DGAPA-UNAM.

\end{document}